\pdfoutput=1
\documentclass[]{aa}
\usepackage[latin9]{inputenc}
\usepackage[active]{srcltx}
\usepackage{amsbsy}
\usepackage{graphicx}
\usepackage{textcomp}
\usepackage[]{natbib}
\bibpunct{(}{)}{;}{a}{}{,}
\makeatletter

\providecommand{\tabularnewline}{\\}

\usepackage{amsbsy}\usepackage{array}
\providecommand{\tabularnewline}{\\}

\begin{document}
\title{Variations of the solar cycle profile in a solar dynamo with fluctuating
dynamo governing parameters }

   \author{V.V.~Pipin\inst{1} \and D.D.~Sokoloff\inst{2} \and I.G.~Usoskin\inst{3,4}}

   \offprints{V.V.Pipin}

   \institute{Institute of Solar-Terrestrial Physics, Russian Academy of
Sciences, Irkutsk, 664033, Russia\and Department of Physics, Moscow
University, 119992 Moscow, Russia \and
 Sodankyl\"a Geophysical Observatory (Oulu Unit), 90014, University
of Oulu, Finland \and
 Dept. of Physical Sciences, 90014 University of Oulu, Finland}

   \date{Received ..... ; accepted .....}

\abstract{Solar cycles vary in their amplitude and shape.
There are several empirical relations between various parameters that
link cycle's shape and amplitude, the foremost the Waldmeier relations.}
{The solar cycle is believed to be a result of the solar dynamo
  action, therefore these relations require an explanation in the
 framework of this theory, which  we aim to present here.}
{We related the cycle-to-cycle variability of solar activity to fluctuations of solar dynamo drivers and
 primarily to fluctuations in the parameter responsible for the recovery of the poloidal magnetic field from
 the toroidal one.
To be specific, we developed a model in the framework of the mean-field dynamo based on the differential rotation and
$\alpha$-effect.}
{We demonstrate that the mean-field dynamo model, which is based on a
  realistic rotation profile and on nonlinearity that is associated with the magnetic
 helicity balance, reproduces both qualitatively and quantitatively the Waldmeier relations observed in sunspot data since 1750.
The model also reproduces more or less successfully other relations between the parameters under discussion,
 in particular, the link between odd and even cycles (Gnevyshev-Ohl rule).}
{We conclude that the contemporary solar dynamo theory provides a way to explain the cycle-to-cycle
variability of solar activity as recorded in sunspots. { We discuss the importance of the model for
stellar activity cycles which, as known from the data of the Mount
Wilson HK project, which measures the Ca H and K line index for other
stars, demonstrate the cycle-to-cycle variability similar to solar cycles.}}

\keywords{Sun: activity -- Sun: magnetic field -- Sun: sunspots --
Stars: activity -- Dynamo}

\titlerunning{Variations of solar cycle profile in a solar dynamo}
\authorrunning{Pipin et al.}

\maketitle

\section{Introduction\label{intr}}

Solar activity has a periodic nature, but the cycle amplitude and
 shape vary from one cycle to the other. { This challenges the
   prognostic abilities of solar activity models.
The sunspot activity can be quantified by using various tracers derived from
observations. These tracers show an interrelation among each
other. The indices characterizing the tracers can be employed to
predict the future evolution of solar activity.}
Waldmeier (\citeyear{w35}) first suggested this
 option (an inverse correlation between the length of the ascending
 phase of a cycle and the peak sunspot number of that cycle) and
 applied it (\citealp{w36}) to predict the subsequent cycle.
Later, other relations of this kind were proposed and
 called Waldmeier relations (see for a review \citealp{vetal86}
 and \citealp{hathaw02}).
The nature of physical processes underlying the Waldmeier relations is not clear
 \citep[see discussion, e.g., in][]{camschu07,detal08,karak11}.
We note,  that these statistical properties of the magnetic activity also exist for
other tracers related to the sunspot activity (e.g., sunspot group
 number and area, see \citealp{vetal86,hathaw02,karak11}),
and even for the other kind of solar and stellar activity
indices, e.g., for the Ca II index \citep{soon94}.
The Waldmeier  relations are considered as a valuable test for dynamo models
\citep{karak11,pk11}.

Clarifying the physics underlying the Waldmeier
relations is particularly attractive to support the prognostic abilities concerning solar cycle.
It is believed that the cyclic solar activity is driven by
a dynamo, i.e. a mechanism that transforms the kinetic energy of
hydrodynamic motions into a magnetic one.
Many modern solar dynamo models \citep[see, e.g.,][]{stix:02}  assume that the
toroidal magnetic field that emerges on the surface and forms
sunspots is generated near the bottom of the convection zone, in the
tachocline or just beneath it in a convection overshoot layer.
This kind of dynamo can be approximated by the Parker dynamo
waves \citep{park93}.
The direction of the dynamo wave propagation
in the framework of the $\alpha\Omega$-dynamo is defined by the
Parker-Yoshimura rule \citep{par55,yosh1975}, according to which  the
wave propagates along iso-surfaces of the angular velocity.
The propagation can be affected by the turbulent transport (associated
with the mean drift of magnetic activity in the turbulent media by
means of the turbulent mechanisms), by the anisotropic turbulent
diffusivity \citep{k02}, and by meridional circulation \citep{yosh1975,choud95}.
An alternative to the Parker's
surface dynamo waves is the distributed dynamo with
subsurface shear \citep[e.g.,][]{b05}, where the dynamo wave propagates
along the radius in the main part of the solar convection zone \citep{k02}.
Near-surface activity is determined by the subsurface shear.
Another popular option is the flux-transport dynamo \citep[e.g.,][]{choud95,dc99}.

In the context of dynamo theory, the Waldmeier relations can be
explained by invoking physical mechanisms of the solar magnetic
field generation and a mechanism that drives variations of the amplitude and shape
of the activity cycle.
For example, \cite{pk11}, hereafter PK11, showed
that variations of the $\alpha$-effect amplitude may explain the
correlation between the cycle rise rate and the cycle amplitude and
other types of the Waldmeier relations as well.
It was suggested \citep{choud92,h93} that the fluctuations of the $\alpha$-effect
(associated with kinetic helicity fluctuations) are likely to be
one of the natural sources of the cycle parameter variations.

In addition to the statistical relations between the cycle parameters within
a separate cycle there are correlations relating the parameters in
subsequent cycles, for example, the odd-even cycle and the last
cycle period-amplitude effects.
These effects are closely related
to the memory of the dynamo processes and to the strength of the
saturation processes, which damp deviations of the cycle
parameters from the reference state characterizing the cycle
attractor \citep{oss-h96a,oss-h96b}.

It was argued \citep{choud92,h93}, that a dozen
percent is a reasonable estimate for the noise component of the $\alpha$-effect.
Previous calculations (see the above cited papers)
showed that a straightforward application of the idea with the vortex
turnover time and the vortex size as the correlation time and length for the
$\alpha$-fluctuations needs fluctuations much stronger than the mean $\alpha$.
On the other hand, the results of direct  numerical
simulations \citep[e.g,][]{bs02} and results of current helicity
(related to $\alpha$) observations in solar active regions
\citep[e.g.,][]{zetal10} {suggest that the correlation time for $\alpha$-fluctuations
 can be comparable to the cycle length and the correlation length comparable
to the extent of the latitudinal belts}.
Using these results, \citet{moss-sok08} and \citet{uetal09}  showed
that an $\alpha$-noise on the order
of few dozen percents is sufficient to explain the Grand minima of solar activity.
The aim of this paper is to examine the result of $\alpha$-fluctuations in the statistical properties of the solar
cycle including the Waldmeier relations and the odd-even cycle effect.

We chose a particular model for the solar cycle in which
$\alpha$-fluctuations are introduced.
Of course, it is impractical to try all available models to learn which one is better to
obtain the relations under discussion, but we select below the model
among a relative wide choice of the models that gave better results in
the preliminary simulations \citep{ps11}.

\section{Basic equations}

\subsection{Dynamo model}

The dynamo model employed in this paper has been described in detail in \cite{pk11,pk11apjl}.
We study the standard mean-field induction
equation in perfectly conductive media:
\[
\frac{\partial\mathbf{B}}{\partial t}=\boldsymbol{\nabla}\times\left(\mathbf{\boldsymbol{\mathcal{E}}+}\mathbf{U}\times\mathbf{B}\right),
\]
 where $\boldsymbol{\mathcal{E}}=\overline{\mathbf{u\times b}}$ is
the mean electromotive force, with $\mathbf{u,\, b}$ being the turbulent
fluctuating velocity and magnetic field, respectively; $\mathbf{U}$
is the mean velocity (differential rotation), and the axisymmetric magnetic
field is
\[
\mathbf{B}=\mathbf{e}_{\phi}B+\nabla\times\frac{A\mathbf{e}_{\phi}}{r\sin\theta},
\]
 where $\theta$ is the polar angle. The expression for the mean electromotive
 force vector $\boldsymbol{\mathcal{E}}$ is given by
 \cite{pi08Gafd}. It is expressed as follows:
\begin{equation}
\mathcal{E}_{i}=\left(\alpha_{ij}+\left(1+\xi_{\gamma}\right)\gamma_{ij}\right)\overline{B}
-\left(1+\xi_{\eta}\right)\eta_{ijk}\nabla_{j}\overline{B}_{k}.\label{eq:EMF-1}
\end{equation}
 Tensor $\alpha_{ij}$ represents the alpha effect. It includes the
hydrodynamic and magnetic helicity contributions,
\begin{equation}
\alpha_{ij}=C_{\alpha}\left(1
+\xi_{\alpha}\right)\psi_{\alpha}(\beta)\sin^{2}\theta\alpha_{ij}^{(H)}+\alpha_{ij}^{(M)},\label{alp2d}
\end{equation}
where the hydrodynamical part of the $\alpha$-effect is defined by $\alpha_{ij}^{(H)}$,
$\xi_{\alpha,\eta,\gamma}$ defines the noise,
$\psi_{\alpha}\left(\beta\right)$ is  the quenching function,
 where $\beta={\displaystyle
  \frac{\left|\overline{B}\right|}{u'\sqrt{4\pi\overline{\rho}}}}$,
$u'$ is the convective RMS velocity. The reader can find
 expressions for the quenching function, $\psi_{\alpha}$ and
$\alpha_{ij}^{(H)}$ in appendix.

Contribution of the small-scale magnetic helicity $\overline{\chi}=\overline{\mathbf{a\cdot}\mathbf{b}}$
($\mathbf{a}$ is a fluctuating vector-potential of the magnetic field)
to the $\alpha$-effect is defined as $\alpha_{ij}^{(M)}=C_{ij}^{(\chi)}\overline{\chi}$
, where the coefficient $C_{ij}^{(\chi)}$ depends on the turbulent
properties { of the medium
and on the parameter characterizing the influence of the Coriolis
force on convection. Expression for $C_{ij}^{(\chi)}$ is the same as in PK11 and  is given in appendix, as well.}
Other parts of Eq.(\ref{eq:EMF-1})
represent the effects of turbulent pumping, $\gamma_{ij}$, and turbulent
diffusion, $\eta_{ijk}$. { We give their expressions in appendix.}

The nonlinear feedback of the large-scale magnetic field to the $\alpha$-effect
is described as a combination of an {}``algebraic'' quenching by the
function $\psi_{\alpha}\left(\beta\right)$,
and a dynamical quenching due to the magnetic helicity conservation
constraint. The magnetic helicity, $\overline{\chi}$ , subject to
a conservation law, is described by the following equation
 \citep{kle-rog99,sub-bra:04}:
\begin{eqnarray}
\frac{\partial\overline{\chi}}{\partial t} & = & -2\left(\boldsymbol{\mathcal{E}\cdot}\overline{\mathbf{B}}\right)-\frac{\overline{\chi}}{R_{\chi}\tau_{c}}+\boldsymbol{\nabla}\cdot\left(\eta_{\chi}\boldsymbol{\nabla}\bar{\chi}\right),\label{eq:hel}
\end{eqnarray}
 where $\tau_{c}$ is a typical convection turnover time. Parameter
$R_{\chi}$ controls the helicity dissipation rate without specifying
the nature of the loss. It seems reasonable that the helicity
dissipation is most efficient in the near surface layers because of
a strong decrease of $\tau_{c}$ toward the surface. The last term in
Eq.(\ref{eq:hel}) describes the diffusive flux of the magnetic helicity
\citep{mitra10}.

We used the solar convection zone model computed by
\cite{stix:02}, in which the mixing-length is defined as
$\ell=\alpha_{MLT}\left|\Lambda^{(p)}\right|^{-1}$, where
$\mathbf{\boldsymbol{\Lambda}}^{(p)}=\boldsymbol{\nabla}\log\overline{p}\,$
is the pressure variation scale, and $\alpha_{MLT}=2$. The turbulent
diffusivity is parametrized in the form,
$\eta_{T}=C_{\eta}\eta_{T}^{(0)}$, where
$\eta_{T}^{(0)}={\displaystyle \frac{u'\ell}{3}}$ is the
characteristic mixing-length turbulent diffusivity, $\ell$ is the
typical correlation length of turbulent flows, and $C_{\eta}$ is a
constant to control the efficiency of the large-scale magnetic field
dragged by the turbulent flows. Currently, this parameter cannot be
introduced  into the mean-field theory in a consistent way.

In this paper we use $C_{\eta}=0.05$. The fairly
low turbulent diffusivity both due to low  $C_{\eta}\ll 1$ and due to
quenching of the turbulent difffusivity coefficient for the fast-rotation regime in the depth of the solar convection zone, where
$\Omega^{*}=2\Omega_0\tau_c\gg 1$, provides the correct value of the cycle period
in the model.
Note that in the fast-rotation regime the turbulent magnetic
diffusivity is dominated by the anisotropic component of the diffusivity
tensor along the rotation axis. This component is growing in the
intermediate range variations of $\Omega^{*}$. Currently, the problem
with $C_{\eta}\ll 1$ has no satisfactory resolution within the dynamo theory.
In the model we used  an analytical fit to
of the differential rotation profile  proposed by
\cite{antia98}. {It was given in our earlier papers (see Fig. 1a in \citealp{pk11,ps11}).}
\begin{figure*}
\includegraphics[width=0.95\textwidth]{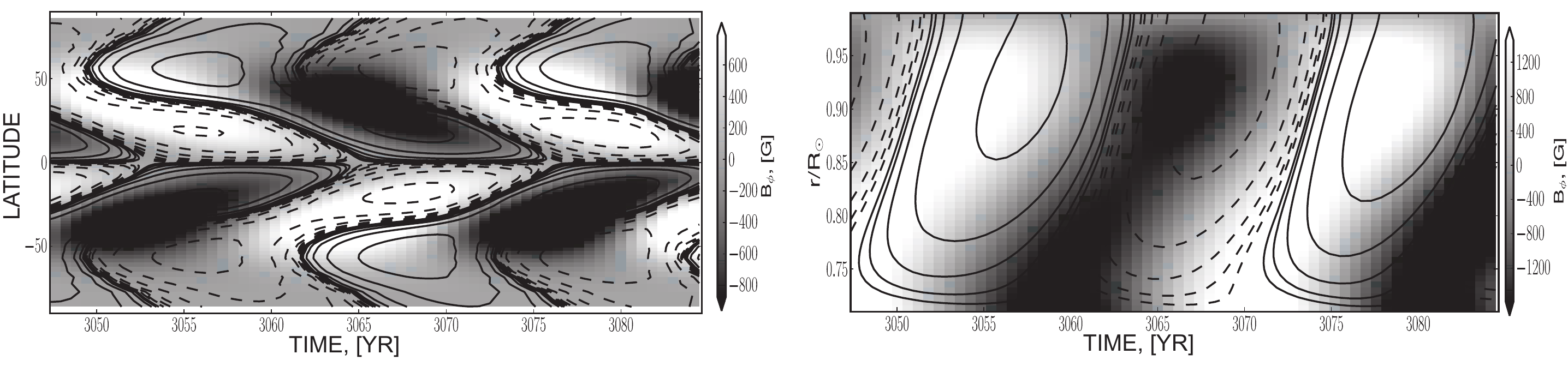}
\caption{\label{2dbat}Typical time-latitude and time-radius
(at the $30^{\circ}$ latitude) diagrams of the toroidal field (gray
scale), the radial field (contours at left panel) and the poloidal
magnetic field (which is drawn by the contours of the
vector-potential at the right panel) evolution in the 2D1$\alpha$ model
(see Table 1). The toroidal field averaged over the subsurface
layers in the range of $0.9-0.99R_{\odot}$, the radial field is
taken at the top of the convection zone. }
\end{figure*}

{ We matched the potential field
outside and the perfect conductivity at the bottom boundary with the standard boundary conditions}.
For magnetic helicity, similar to \cite{guero10} 
we put $\boldsymbol{\nabla_{r}}\bar{\chi}=0$ at the bottom of the domain
and $\bar{\chi}=0$ at the top of the convection zone.
\begin{table}
\caption{\label{modpar} Parameters of the dynamo models}
\begin{centering}
\begin{tabular}{c c c c c c c c}
\hline
Model  & $\eta_{\chi}/\eta_{T}$  &$C_{\alpha}$/$R_{\chi}$  & $B_{0}$  & noise & $\sigma$ \tabularnewline
\hline
2D1$\alpha$   & $10^{-5}$  & 0.03/200  & 800  & $\xi_{\alpha}$ & 0.2\tabularnewline
2D1$\eta$     & -/-        & -/-  &  -/- & $\xi_{\eta}$ & 0.04\tabularnewline
2D1$\alpha$L   & -/-        & -/-  &  -/- & $\exp(\xi_{\alpha})$& $\log\sigma(\xi_{\alpha})$\tabularnewline
\hline
\end{tabular}
\par\end{centering}
\end{table}

The parameters of the model are summarized in the Table \ref{modpar},
{ where
 $\eta_{\chi}/\eta_{T}$ is
the ratio between the turbulent magnetic helicity diffusivity and
the turbulent magnetic diffusivities;  the
parameter $R_{\chi}$ controls the helicity dissipation rate;  $B_{0}$ is a typical
strength of the toroidal magnetic field controlling the sunspots
number parameter in the 2D models; the column ``noise'' defines the
fluctuating parameter and $\sigma$ is the standard deviation of the Gaussian noise in the
model. The lognormal noise in the model 2D1$\alpha$L was symbolically
denoted as $\exp(\xi_{\alpha})$.}

The  left panel in Figure \ref{2dbat} shows a typical time-latitude
diagram in the model 2D1$\alpha$ for the toroidal magnetic field evolution averaged over the subsurface layers
$0.9-0.99R_{\odot}$ and the radial magnetic at the top of the integration domain.
The right panel shows the time-radius diagram for the toroidal an
poloidal components of the large-scale magnetic
 field evolution at $30^{\circ}$ latitude. { Note that the geometry of
 the poloidal magnetic field is drawn by iso-contours of the toroidal vector-potential.}

\subsection{Noise model}
\begin{figure}[t]
\includegraphics[width=0.49\columnwidth]{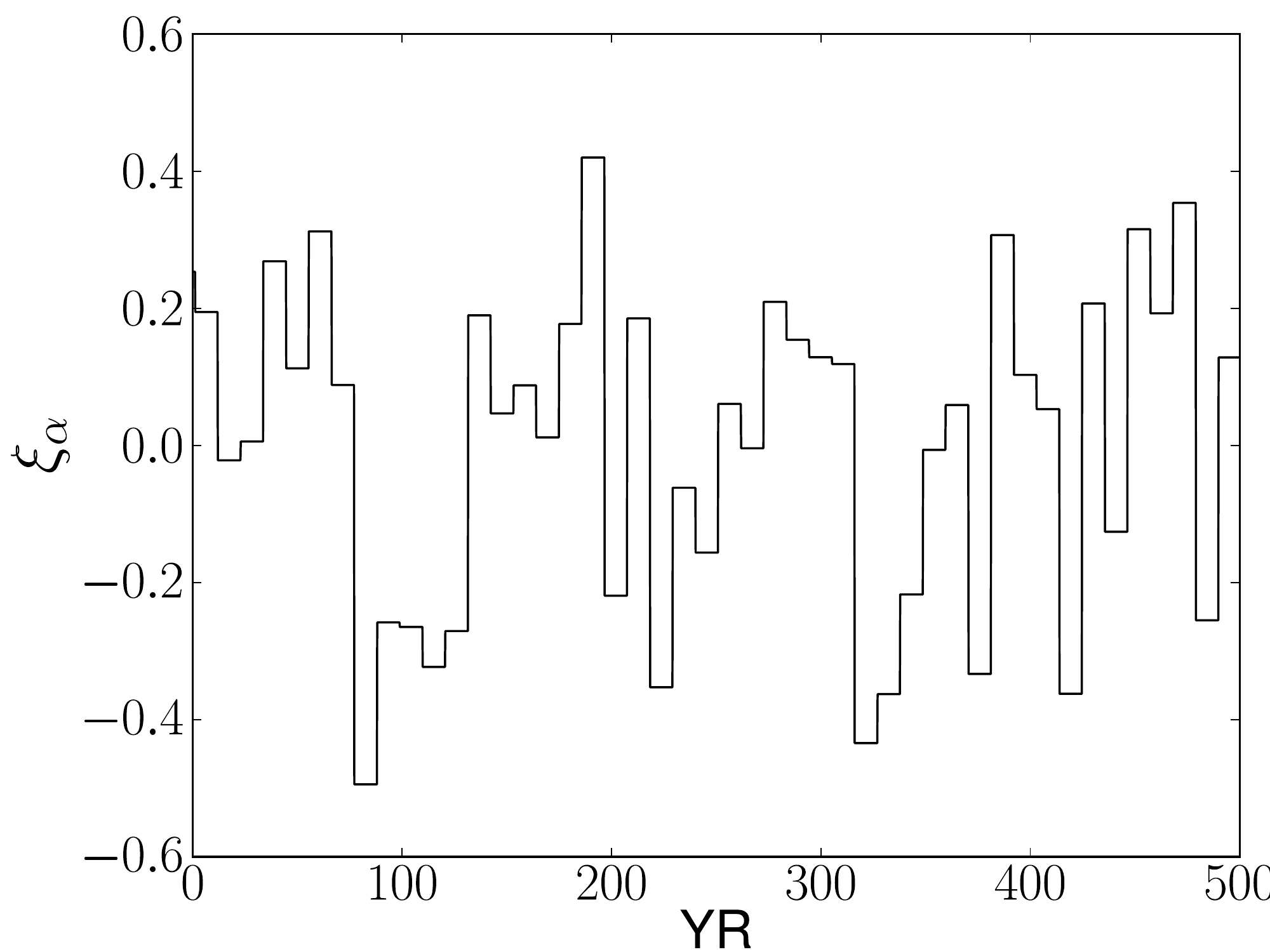}
\includegraphics[width=0.49\columnwidth]{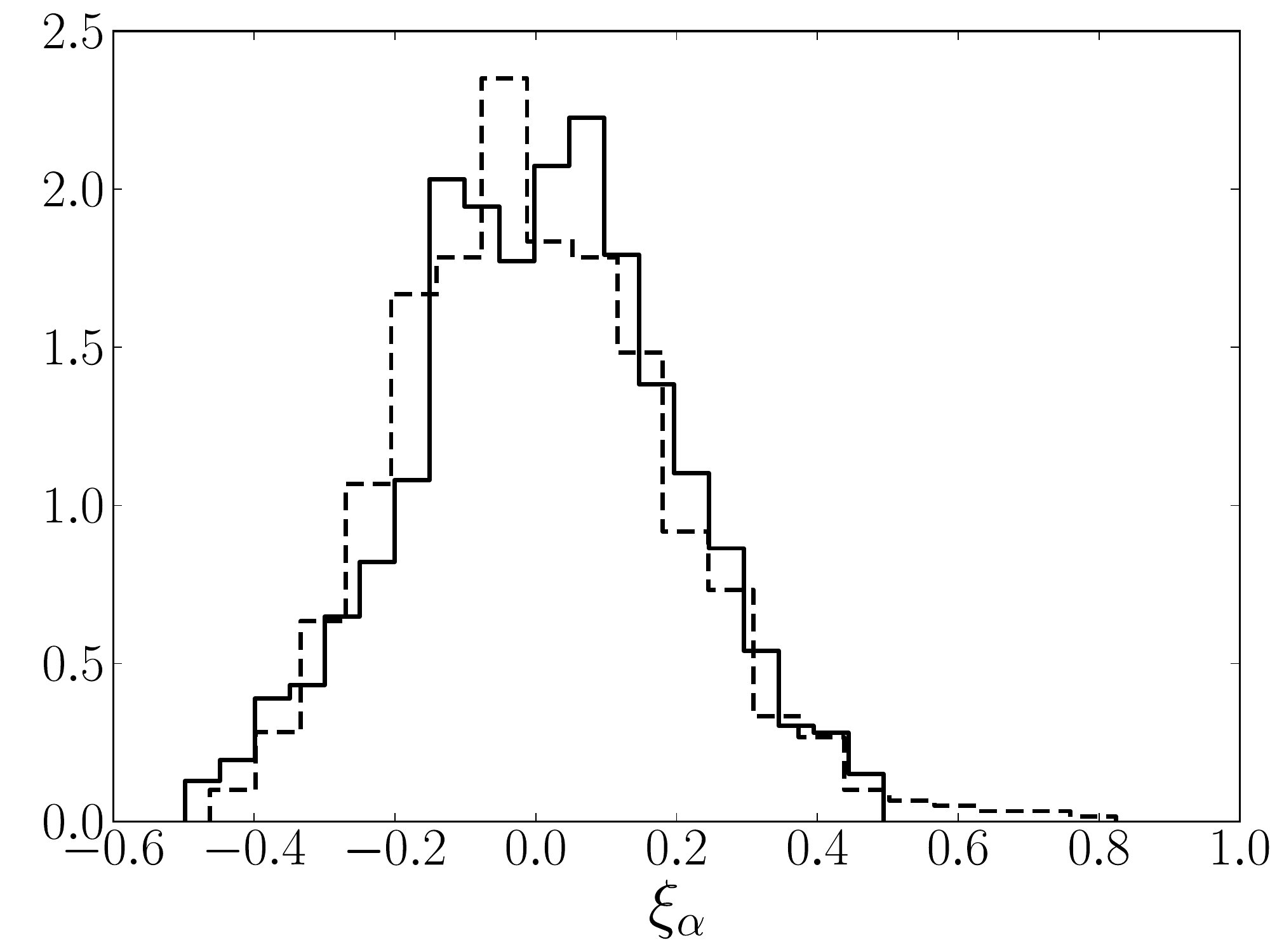}
\caption{\label{nois-hyst} Typical realization of the fluctuating
  part of the $\alpha$-effect (left panel) and its probability distribution
  function (solid line, right panel). There we show the PDF for the
  lognormal fluctuations as well (dashed line).}
\end{figure}
In Eq.(\ref{eq:EMF-1}), the noise, $\xi_{\alpha,\eta,\gamma}$, contributes to the hydrodynamic part
of the $\alpha$-effect (see, Eq.(\ref{alp2d})), to the turbulent
diffusion, and to the turbulent pumping. Following \cite{uetal09} the model employs the long-term
Gaussian fluctuating $\xi_{\alpha,\eta}$ of the low amplitude with
RMS deviation given in the Table 1 (last column). It is expected
from general consideration that, for the low amplitude
fluctuations, $\xi_{\eta}$ is an order of magnitude smaller than
$\xi_{\alpha}$ (because $\alpha^{(H)}\sim u'$ and $\eta_T\sim
u'^2$). To examine the long-term dynamics of the model with regard to
the specific statistical distribution of the noise we included
the results for a model with the lognormal distribution of
$\xi_{\alpha}$ (see, Subsection \ref{LT}). In this case the
parameters of the lognormal distribution were computed from the
corresponding Gaussian distribution.
{
{ Random numbers were generated using the Numerical Recipes Fortran
 subroutine {\it ``gasdev''}.} The
amplitudes of the fluctuations were restricted to
2$\sigma$. A realization of the
lognormal fluctuations was prepared before the run. We took the input
parameters from the realization of the Gaussian distribution fluctuations,
which were computed with taking into account the 2$\sigma$ cut-off.
The fluctuation renewal time was constant and equal to the period of the cycle in the
model. Figure \ref{nois-hyst} shows a typical realization for the Gaussian fluctuations of the
$\alpha$-effect, its probability distribution function (PDF), and the PDF
of the generated lognormal fluctuations.}

{ We considered the amplitude of the
standard deviation of the $\alpha$-coefficient as an input
parameter. We chose this quite arbitrary based on the following
crude estimation. The total number of cyclones in the solar convection
zone  may be of the order of $N = 2 \times 10^3$, see, e.g., \cite{miesch08},
who found that the solar convection vertical vorticity spatial spectrum 
is flat and has a maximum
at about $\ell\approx 140$, where  $\ell$ is the spatial wavenumber. If
$\alpha \approx 0.1 u'_c$ then the $1/\sqrt N$ fraction of the magnitude fluctuations in velocity
corresponds to about 20\% of the fluctuations in $\alpha$. We stress that
this very crude estimate needs a support from detailed numerical
modeling and observations, which are, however, obviously out of the
scope of this paper. The Gaussian fluctuations were cut off on the
$2\sigma$-level to exclude a possible effect of rare and very strong
fluctuations, which are hardly associated with Waldmeier relations.
It can be important in principle for Grand minima statistics. 
 We did not address the latter possibility systematically,  but tested the
lognormal fluctuations to study the impact of the PDF tail on the
long-term solar cycle variations.}

\subsection{The sunspot cycle model and the Waldmeier relations\label{ssndef}}

Here, we define the Waldmeier relations as a set of the
mean properties of the sunspot cycle including relations between
the rise rate of the cycle and its amplitude and the relation
``period-amplitude''.
In the original form the Waldmeier
relation reads as a link between the period of a cycle and amplitude
of the subsequent cycle. Other relations like this (rise time and
amplitude; rise time and decay time) are sometimes referred to
Waldmeier relations,  as well.
 These relations were considered in PK11 and
in our previous paper \citep{ps11}. The ratio of the decay
and rise rates is known as the shape of the cycle.
Amother relation presented in literature is the so-called Gnevyshev-Ohl
rule \citep[e.g.][]{chetal07}, which provides a positive correlation
between the amplitudes or intensities of $2N$th and $2N+1$th cycles.

The amplitude of a cycle is defined as the difference between the
maximum sunspot number and the sunspot number in the preceding
minimum. The latter can differ from zero because of the overlap of
subsequent cycles. { The cycle period is equal to the time between the
subsequent minima. The rise time of a cycle is defined by the
difference between the moment of the cycle maximum and the moment of
the preceding minimum.} The rise rate is defined as the ratio between
the difference of the sunspot numbers at maximum and minimum of the
cycle and the rise time of the cycle. There is similar definition for
the decay rate of the cycle.

{ Following to PK11 we relate the sunspot number with the toroidal
magnetic fields in the near-surface layer.} {In PK11 the instant
sunspot number was defined  using the maximum of the toroidal
magnetic field strength, which was taken over all latitudes and
averaged over the surface layer from 0.9 to 0.99$R_{\odot}$. Then, this
value was related to the sunspot number (SSN) via the three-halves law
 proposed by Bracewell (1988) to relate SSN in the cycle with
a ``linear'' sinusoidal part of the SSN variation. The relation between
the toroidal field and SSN, which was introduced in PK11 has
an undesirable property of giving non-zero SSN at the minimum of a
cycle and strong variations of the minima amplitudes with
variations of the $\alpha$-effect parameters (see Fig.6 there).}

Here, we examined another possibility (also see \citealp{ps11}).
We assumed that sunspots are produced from
the toroidal magnetic fields by means of the nonlinear instability,
and avoided to consider the instability in details. To model the sunspot
number $W$ produced by the dynamo we used the following ansatz:
\begin{equation}
W\left(t\right)=C_{W}\left\langle B_{{\rm max}}\right\rangle _{SL}
\exp\left(-\frac{B_{0}}{\left\langle B_{{\rm max}}\right\rangle _{SL}}\right),\label{eq:wolf}
\end{equation}
 where  $\left\langle B_{{\rm max}}\right\rangle _{SL}$
is the maximum of the toroidal magnetic field strength over latitudes
averaged over the subsurface layers in the range of
$0.9-0.99R_{\odot}$;
 $B_{0}$ is a typical strength of the toroidal magnetic
field sufficient to produce the sunspot, hereafter we put
$B_0=800$G;  $C_{W}$ is the parameter
to calibrate the modeled sunspot number relative to observations.
Hereafter we put $C_{W}=1$. {Results by \cite{ps11} showed that similar
to PK11 the Waldmeier relations can be reproduced with the Wolf number
definition (Eq.\ref{eq:wolf}). The exponential dependence in
Eq.(\ref{eq:wolf}) yields the Waldmeier relations at smaller variations
of the $\alpha$-effect compared to those in PK11, where the Bracewell
law was used. Also, we find that the mean-field
 dynamo model with relation (\ref{eq:wolf})  reproduces the
long-term variation of the cycle  much better than in the case of  the
three-halves law.}

\subsection{Observational data set}

Although the series of group sunspot numbers covers 400 years since
1610 AD \citep{hs98}, giving a measure of the temporal variability of
solar activity, parameters of the solar cycle such as its total
length and ascending/descending phases are not reliably known for
the earlier times. Solar cycle parameters can be more or less
reliably evaluated since 1750 or, with some caveats, after the end
of the Maunder minimum in 1712 \citep{ius08}. However, even in this
case an uncertainty related to the potentially lost solar cycle in
the last decade of 18-century \citep[e.g.][]{ius03,ius09} still exists.
The Sun was
amazingly well observed during the Maunder minimum, especially in
its second half \citep{ribes93}, but the solar cycle was suppressed
below the threshold for sunspot formation, which led to unclear dynamo
manifestations. Cycles before the Maunder minimum are not well known
\citep{vaq11} and their shapes cannot be obtained. Therefore, only
the period of 1750-2009 AD, which includes 23 full solar cycles in the
official numeration, can be analyzed here.
Statistical
properties of the long-term variations of the solar cycle can be
estimated on the base of the reconstructed data set proposed by
\cite{uetal04} and \cite{setal04}.

There are several synthetic series that present solar cyclic
variability for the times before the beginning of the sunspot
series. They are based on a fit of a prescribed mathematical model
to fragmentary non-systematic qualitative proxy data of naked-eye
sunspot or auroral observation \cite[e.g.,][]{shov55,n97}.
These synthetic series do not pretend to be quantitative
reconstructions of solar activity and cannot be used to
analyze of solar cycle parameters, which are explicitly prescribed
in the model rather than reconstructed.

Although sunspot activity is greatly suppressed during Grand minima,
the solar dynamo continues to operate at a reduced level. For example, an
analysis of sunspot and aurora \citep{kp88} data during the Maunder
minimum suggests that the dominant periodicity was shifted from the
11 years to ~20-22 years \citep{s92,ius01}. Data of the
cosmogenic isotope $^{14}$C also confirm longer cycles during the
Maunder minimum \citep{pd98,miy06a}. A similar
lengthening of the solar cycle during a Grand minimum has been
observed, using the $^{14}$C data, also for the Sp\"orer minimum at the
turn of  the 15-16-th centuries \citep{miy06b}. However, the parameters of
individual solar cycles cannot be recovered for the Grand minima
periods, only the statistical features.

Taking into account all the above information, we  compared
simulations with the monthly smoothed sunspot number (SSN) data set
from the SIDC (Solar Influence Data Center),
which starts at 1750. The data set was additionally smoothed by
means of the Wiener filter. To compute the wavelet spectra
of the sunspot data set, we used the data set provided by
\cite{hs98} and \cite{setal04}.

\section{Results}

We performed long-term simulations for the time interval of about
$10^4$ years (i.e. the time-span  of the longest reconstruction of
the solar cycle history \citep{setal04} using our basic model. To
compare  the results obtained with other dynamo models see
\cite{ps11}. The simulated time data set for $W(t)$ is shown in
Fig.~\ref{waldr} (top panel). It shows events with the extended
period of the high- and low magnetic activity.

\subsection{The Waldmeier relations and the odd-even cycle effect}

\begin{figure*}
\begin{centering}
\includegraphics[width=0.8\paperwidth]{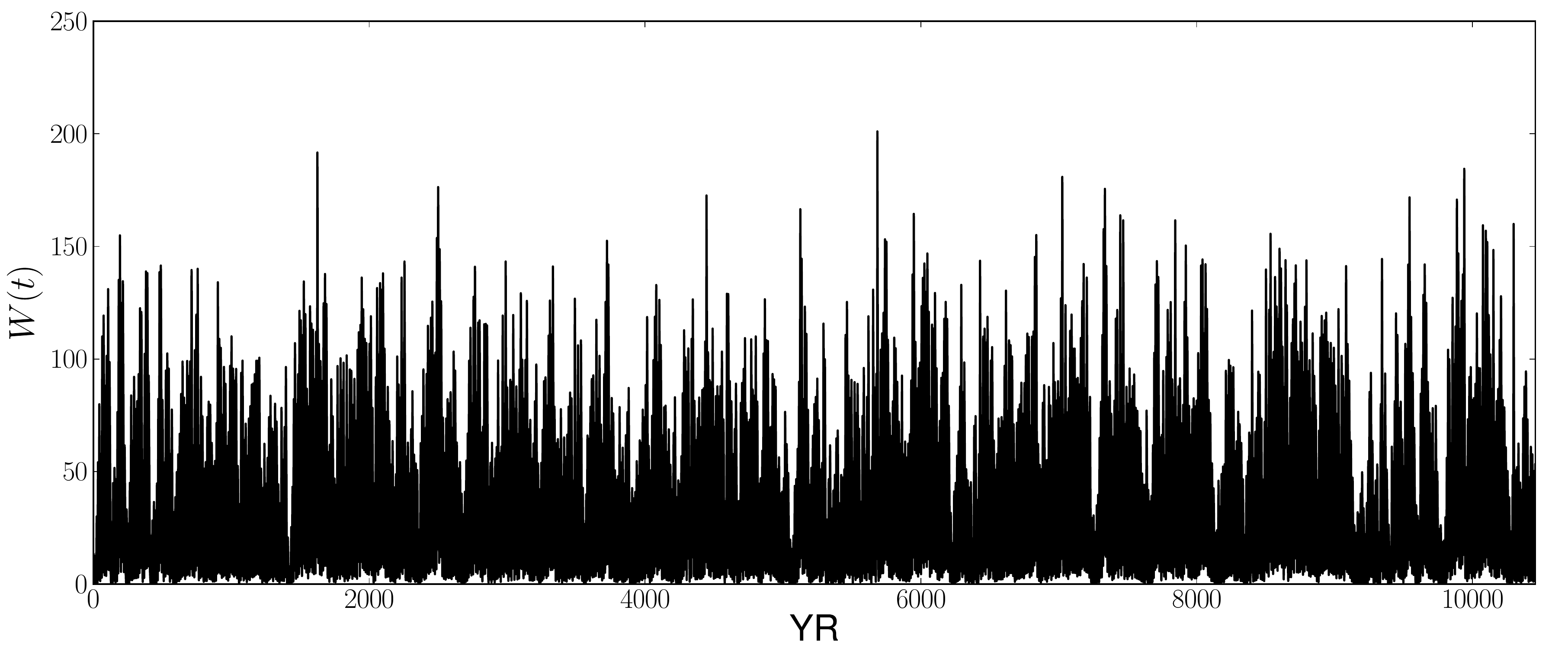}
\includegraphics[width=0.8\paperwidth]{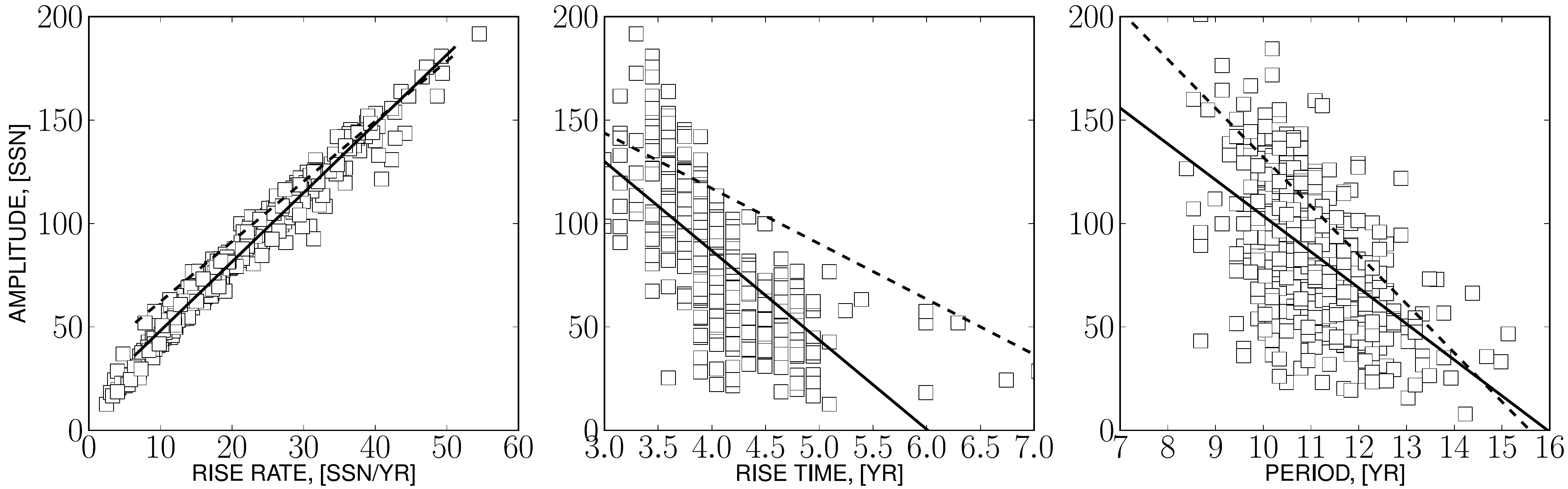}
\par\end{centering}
\caption{Top, the simulated $W(t)$ in our dynamo model.\label{waldr}
Bottom, the Waldmeier relations for the model; squares show data for
individual cycles, while the solid line gives the correlation, the dashed
line shows these relations obtained from the actual SSN data. }
\end{figure*}

First of all, we divided the set into separate cycles to
look for the Waldmeier relations (rise rate to amplitude and
period to amplitude) and the odd-even cycle effect (Gnevyshev-Ohl
rule).
The results are shown in the bottom panel of
Fig.~\ref{waldr}, which depicts spreads of simulated points over
corresponding planes, and their linear fits (solid lines).
The dashed lines show the correlations computed on the base of the
 actual sunspot data.
The results for the rise rate to decay rate relation is
similar to that for the rise rate to amplitude relation and
can be found in \cite{ps11}. Data concerning the linear fits shown
in Fig.~\ref{waldr} are given in Table~\ref{FIT}, { where
 the first four rows contain information for the mean and
variance (standard deviation) for the parameters of the sunspot
cycles in the different data sets. The shape of the cycle is defined
as the ratio between the decay rate and the rise rate of the
cycle. The last
five rows show  linear fits with the mean-square error bar and
the correlation coefficient for the Waldmeier relations and for the
odd-even cycle effect, (I) marks the effect that is calculated on
the base of the SN integrated over the cycle and (II) marks the
effect that is calculated on the base of the cycle amplitudes.}.

{ We see from the Fig.~\ref{waldr} that the model reproduces the
Waldmeier relations  and the Gnevyshev-Ohl rule
reasonably}. Note that the dispersion of
both the simulated and observational data from the linear fit of the rise
rate to amplitude (as well as that one for rise rate to decay rate
\cite{ps11}) are much lower than that for the the period to amplitude
relation and Gnevyshev-Ohl rule. We composed (see Fig. 2, bottom
right) a relation of the rise time to amplitude (i.e. using the quantity
inverse to the rise rate) to learn that the dispersion looks more or
less like that for the relation period-amplitude. As discussed by
\cite{camschu07}, a relation rise time to amplitude has higher
dispersion than that for the rise rate to amplitude. We conclude that
the quality of fitting substantially depends  on the presentation
method chosen to illustrate a relation.

\begin{figure*}
\begin{centering}
\includegraphics[width=0.6\paperwidth]{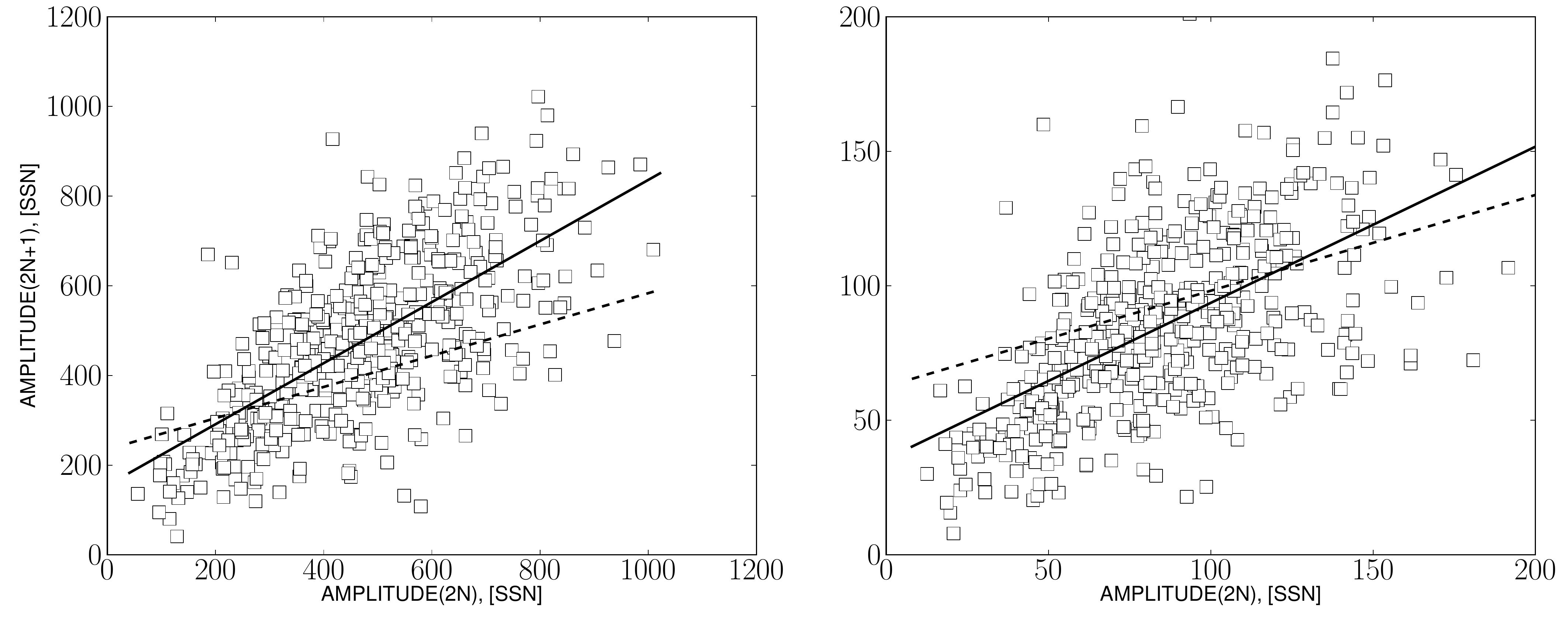}
\par\end{centering}
\caption{Odd-even cycle effect (Gnevyshev-Ohl rule): the left panel
shows the effect that is calibrated on the base of the cycle
amplitude, the right panel shows the effect that is calibrated on the
basis of the SN integrated over the cycle. Squares show data for
individual cycles, while the solid line gives the correlation, the dashed
line shows these relations obtained from the actual SSN data. } \label{GOR}
\end{figure*}

The results concerning the even-odd effect are shown in
Fig~\ref{GOR}. Note that the original formulation of the
Gnevyshev-Ohl rule is calculated  on the basis of the sunspot number (SN) integrated
over the cycle (Fig.~\ref{GOR} right), while the left panel of the
figure shows the effect calculated for the cycle amplitude. We see
from the figure that the model reproduces observations in both cases more or less reasonable,
 however, the slope substantially depends on the definition of the Gnevyshev-Ohl rule.

\begin{table}

\caption{\label{FIT}The parameters of the cycle given by the dynamo medel 2D1$\alpha$ and
  by the monthly smoothed actual sunspot number (SSN).}
\begin{tabular}{ >{\centering}p{2.2cm} > {\centering}p{2.5cm} >{\centering}p{2.5cm} }
\hline
 & 2D1$\alpha$ & SSN \tabularnewline\hline
Period & 11.07$\pm$1.08 & 11.01$\pm$1.12 \tabularnewline\hline
Amplitude & 103.3$\pm$40.5 & 108.2$\pm$38.1 \tabularnewline \hline
Rise Time & 4.06$\pm$.77 & 4.32$\pm$1.07 \tabularnewline \hline
Shape & .59$\pm$0.15 & .69$\pm$0.31 \tabularnewline \hline
Rise Rate - Amplitude & 3.3x+14.8$\pm$6,

0.98$\pm$ 0.001 & 2.9x+33.2$\pm$8.9,

0.97$\pm$ 0.01 \tabularnewline \hline
Period - Amplitude & -17.4x+277.6

$\pm$27.5,

-0.54$\pm$ 0.02 & -23.6x+368.5

$\pm$28.0,

-0.68$\pm$ 0.12\tabularnewline \hline
Rise Time - Amplitude & -43.1x+259.4

$\pm$24.2,

-0.67$\pm$ 0.02 & -26.7x+224.

$\pm$25.,

-0.76$\pm$ 0.1 \tabularnewline \hline
Odd - Even(I) & 0.68x+155.$\pm$136.,

0.67$\pm$ 0.03 & 0.35x+235$\pm$145,

0.33 $\pm$ 0.3 \tabularnewline \hline
Odd - Even(II) & 0.58x+35.6$\pm$26.6,

0.58$\pm$ 0.03 & 0.35x+62.5$\pm$32.3,

0.42$\pm$ 0.28 \tabularnewline \hline

\end{tabular}

\end{table}

It is expected that the strength of the sunspot cycle depends on the
strength of the poloidal field of the Sun in the preceding solar
minimum. Following this idea, \cite{scetal78} suggested using the
strength of the Sun's polar field for the cycle prediction.
 Recently, this idea was exploited
in the Babcock-Leighton type model, see the review by \cite{hath09}.

\begin{figure*}
\begin{centering}
\includegraphics[width=0.8\paperwidth]{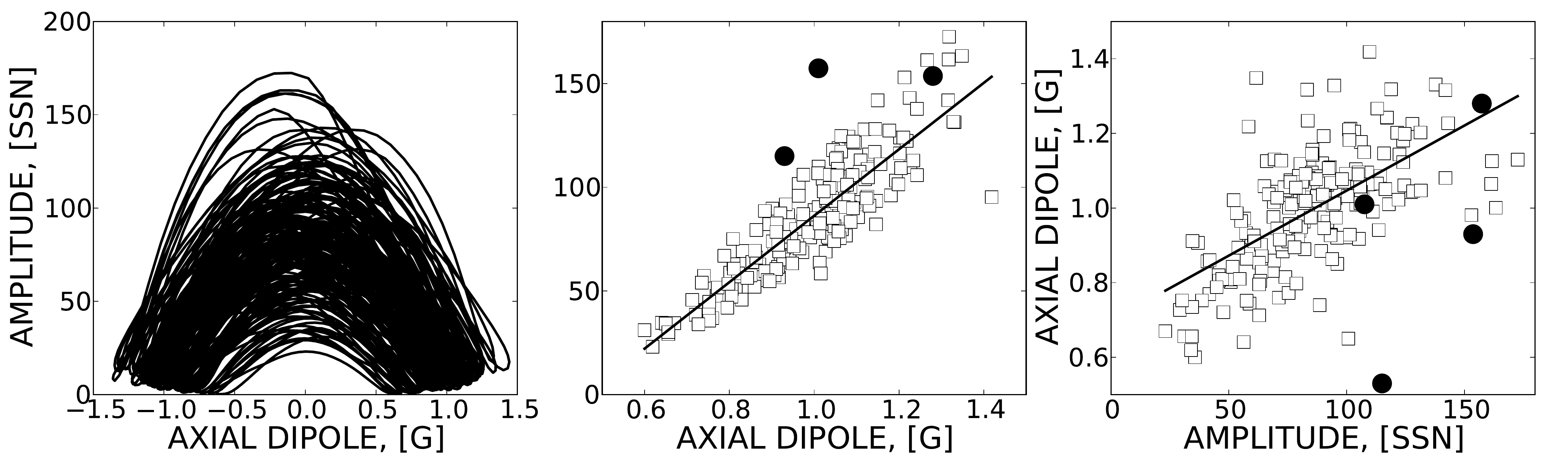}
\par\end{centering}
\caption{Left, the phase relation between the cycle amplitude and the
  strength of the dipole component of the dynamo generated magnetic
  field at the surface. Middle, correlation between dipole component
  at the cycle minimum and the amplitude of the subsequent
  cycle. Right, correlation between the amplitude of the
  cycle  and the  dipole component
  at the  subsequent cycle minimum. The circles mark the results of the
  WSO observations and the SSN data.} \label{minsurf}
\end{figure*}

Figure~\ref{minsurf} illustrates the phase relation between the
amplitude of the sunspot cycle and the strength of the dipole component of the dynamo-generated
magnetic field. There we show the backward and forward correlation between these parameters
of the model, taking the strength of the dipole component at the
cycle minimum. The strength of the dipole component refers to the
surface, and it was calculated as the
first coefficient in the spectral decomposition of the magnetic potential
$A$. The backward correlation has the correlation coefficient
$0.86\pm0.01$ and approximation
$145.4x-74.\pm 13$. The forward correlation between the cycle amplitude
and the strength of the dipole component of the dynamo-generated
magnetic field at the subsequent minimum has the correlation
coefficient $0.64\pm0.01$ and approximation
 $0.004 x + 0.8 \pm 0.12$.
This relation has a higher dispersion than the results 
the fluctuation of the $\alpha$-effect.  For comparison we added several
points obtained from the WSO polar magnetic field observations \citep{setal78,hoek95}

\subsection{Other perspectives}

The main aim of this paper is to demonstrate that fluctuations of
the $\alpha$-coefficient provide an option to explain short-term
dynamics of solar activity cycle such as Waldmeier and similar
relations. We note, on one hand, that this idea can be useful to
explain more long-term dynamics and, on the other hand, that
$\alpha$ is far from being a unique transport coefficient in dynamo
equations, which can be noisy. These noisy transport coefficients can
obviously contribute the activity cycle dynamics. 
Of course, a detailed investigation of these options
is far beyond the scope of this paper, but we present below some
exploratory results in these directions.

\subsubsection{$\eta$-fluctuations vs $\alpha$-fluctuations}

Obviously, fluctuations of the $\alpha$-coefficient affect the solar
activity evolution together with fluctuations of other dynamo
governing parameters. Comparing the relative role of various
fluctuations in the solar cycle variations is beyond the scope of
this paper and we restrict presentation by comparison of the effect
of $\eta$- and $\alpha$-fluctuations only.

\begin{figure}
\begin{centering}
\includegraphics[width=0.5\columnwidth]{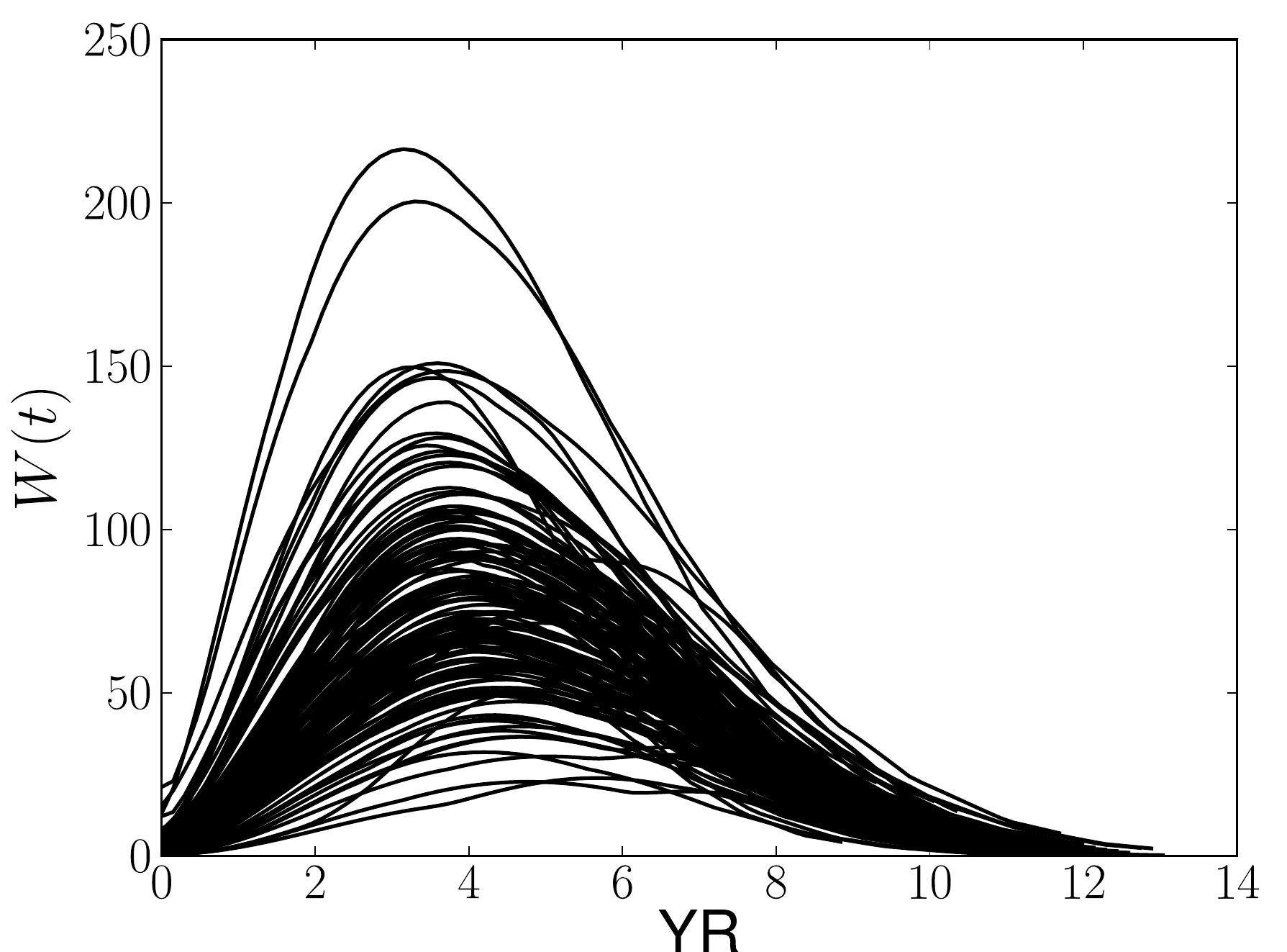}\includegraphics[width=0.5\columnwidth]{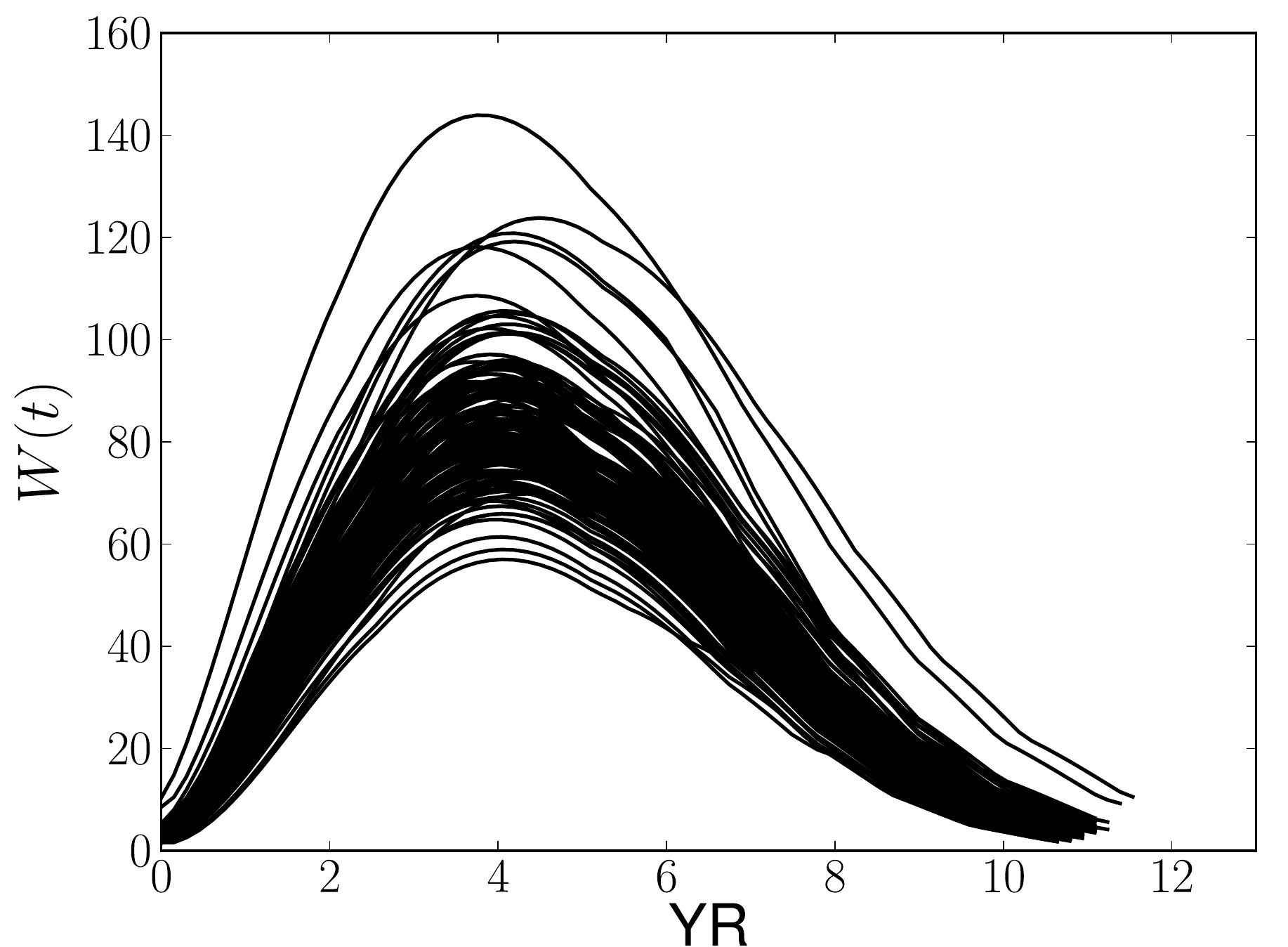}
\par\end{centering}

\caption{Cycles distribution for the 1500yr data set from 2D1$\alpha$
model ($\xi_{\alpha}$-noise), left and 2D1$\eta$ model
($\xi_{\eta}$-noise), right. \label{compar}}
\end{figure}

\cite{choud92}  addressed this point and suggested the following
relation for the fractional growth rate $\Gamma$ of perturbations in
the dynamo as a result of the perturbation of the $\alpha$-effect and
the turbulent diffusion (in our notations):
\[
\Gamma
=\frac{P}{T_{D}}\left(-\xi_{\eta}+\frac{P^{2}}{T_{D}^{2}+P^{2}}\xi_{\alpha}\right),
\]
where $P$ is the period of the cycle and $T_{D}$ is the typical
diffusive time of the dynamo. This relation gives a hint that the
fluctuations of the turbulent diffusivity are relatively more
significant for the dynamo perturbation provided the values
$\xi_\alpha/\alpha$ and $\xi_eta/\eta$ are comparable. The point is,
however, that $\alpha \sim u$ while $\eta \sim u^2$, where $u$ is
the turbulent velocity. We performed simulations and present their
results in Fig.~\ref{compar} for our model with
$\alpha$-fluctuations of 20\% (Fig.~\ref{compar}, left)  and 4\%
$\eta$-fluctuations (Fig.~\ref{compar} right), which corresponds to
the comparable fluctuations in $u$. We see in Fig.~\ref{compar} that the
$\alpha$-fluctuation looks more pronounced than
$\eta$-fluctuations.

{ We examined the influence of the pumping effect fluctuations  on
the cycle variations. The parameters of the model are the same as for
2D1$\alpha$, except that we let the $\alpha$-effect be constant and
$\xi_{\gamma}$ has the same characteristics as $\xi_{\alpha}$ in
2D1$\alpha$. It is found that for a model with $\xi_{\gamma}$ fluctuations the cycle variations look
very similar to the model 2D1$\eta$. The model demonstrates that the
cycle amplitude variations are approximately half of those for the 2D1$\alpha$ model.
Variations of the period are also quite small, like in the model  2D1$\eta$.}

\subsubsection{Resonance effects}

\begin{figure}
\begin{centering}
\includegraphics[width=0.9\columnwidth]{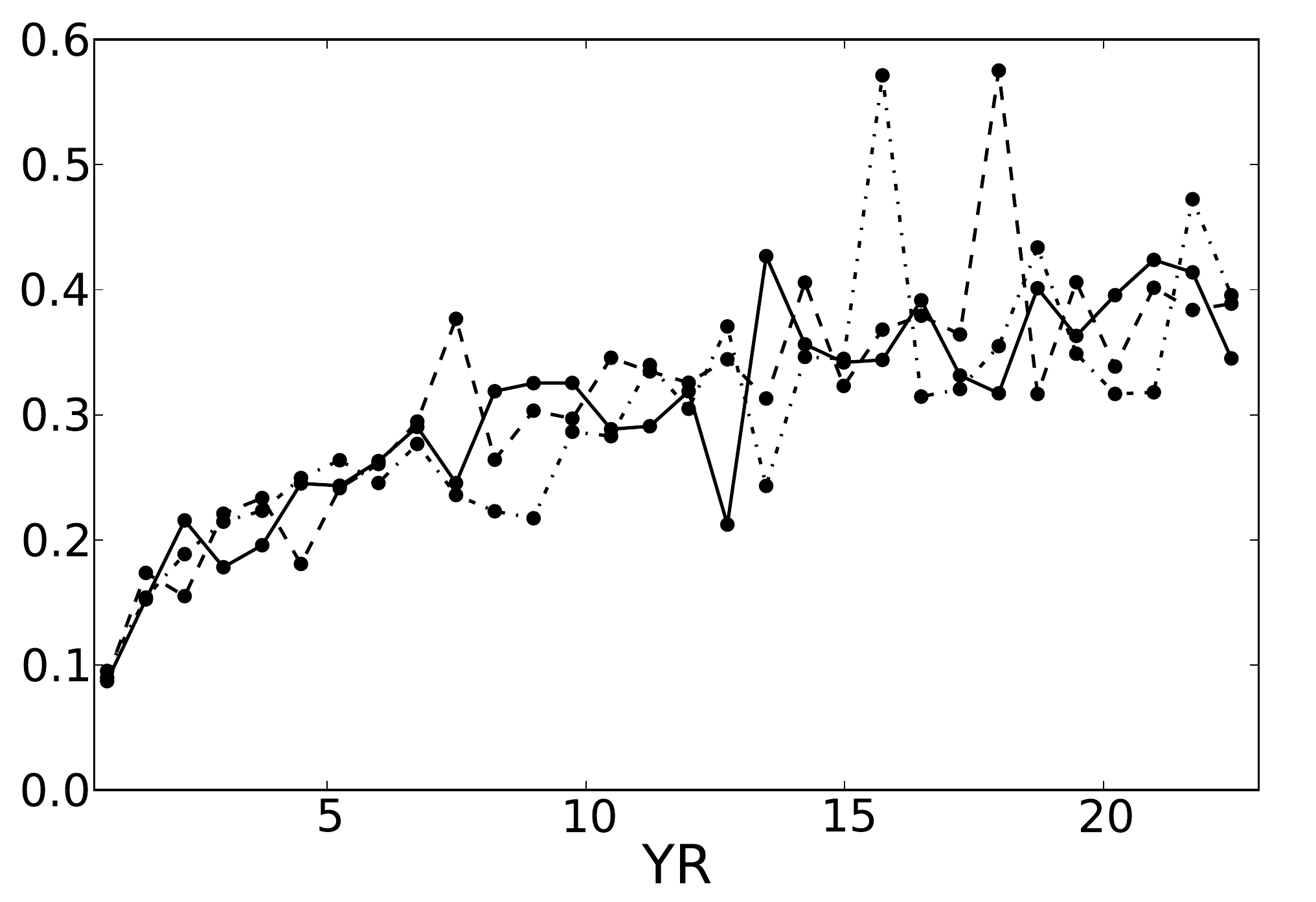}
\caption{Dependence of the cycle amplitude on the noise renewal
time. Three runs with various renovation times are shown with solid,
long-dashed, and short-dashed lines respectively. \label{entropy}}
\par\end{centering}

\end{figure}

Our base model exploits the memory time of $\alpha$-fluctuations
equal to the nominal cycle length. A natural worry is that a
resonance could participate in the Waldmeier relations simulated
while the correlation time of the $\alpha$-coefficient in solar
convection can be different from the cycle length, thus avoiding
resonances. Note that the resonance effects for dynamo waves is
almost not addressed in scientific literature \citep{gd11}. We
varied the correlation time and calculated the cycle amplitude variance
(Fig.~\ref{entropy}). Some peaks are visible in this figure, which
may indicate resonance effects, however, the renovation
times to which they are attached vary from one run to the next.
On the other hand, the results given in
Figure~\ref{minsurf}(right) suggest
that the  resonance effects may depend on the phase
synchronization between the fluctuations of the $\alpha$-effect and the
cycle variations. Therefore, we may  anticipate that the fluctuations on the
descending phase of the solar cycle are more effective than those
on the rise phase of the sunspot cycle. Bearing in mind the
distributed character of the dynamo model, we
 conclude that the resonance phenomena that possibly play a role
here need to be addressed separately \citep[cf.][]{gd11}.

\subsubsection{Long-term dynamics \label{LT}}

We move from the dynamics activity cycles, viz. the timescale of
several dozens cycles where the Waldmeier relations  and the
even-odd relation is applicable, to  the much longer term history of the
solar cycle on timescales of up to $10^4$ years. Here we cannot
discuss such fine details as the Waldmeier relations because the
available data does not trace the cycle shape. The available
isotopic data \citep[we refer to the reconstruction by][]{setal04}
only traces the evolution of the cycle-averaged quantities. Because the
present dynamo model was adopted (via tuning the parameters
$C_{\alpha},R_{\chi}$ and $\eta_{\chi}$) to reproduce {
short-term dynamics, we expect that the long-term dynamics will be
reproduced not as well.}

\begin{figure*}
\begin{centering}
\includegraphics[width=0.4\paperwidth]{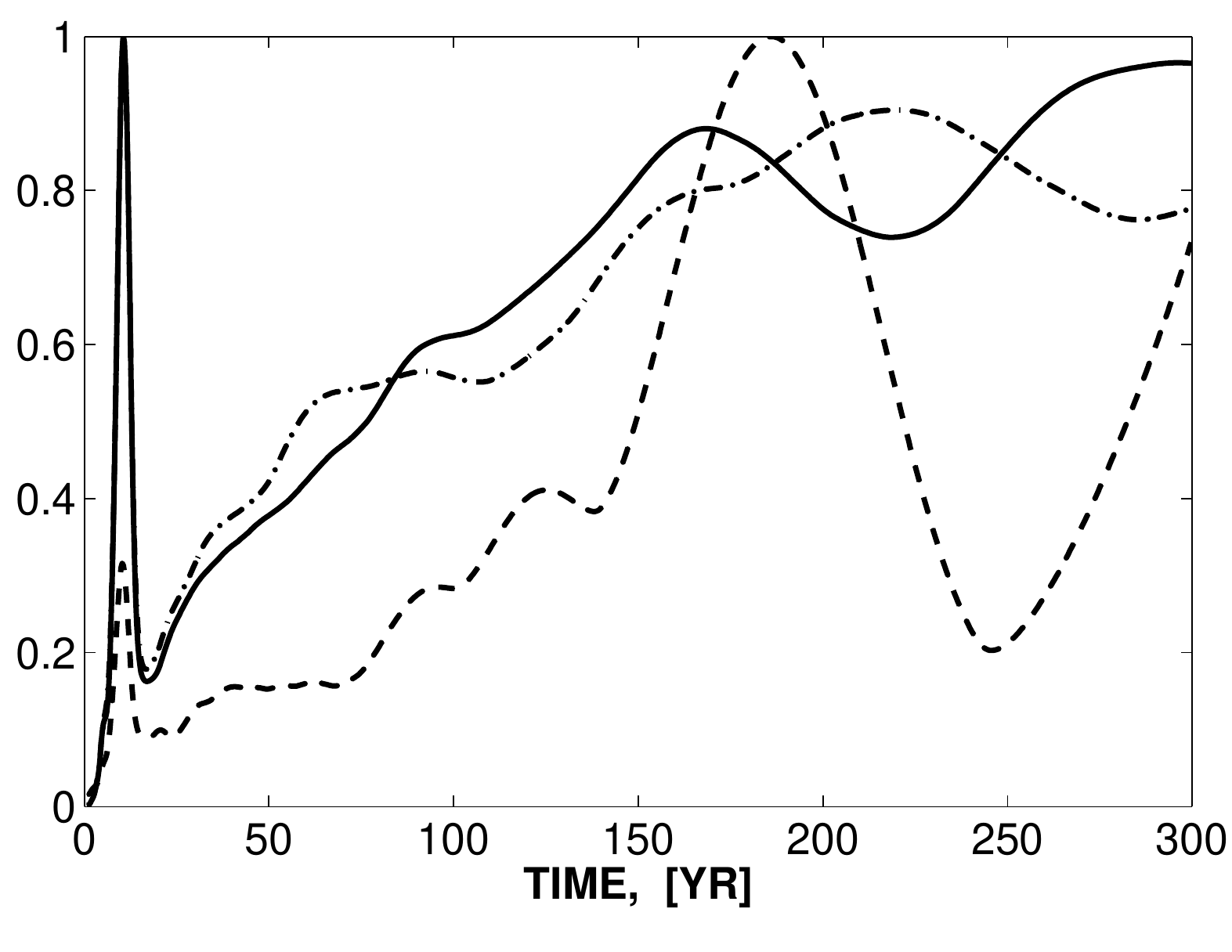}
\includegraphics[width=0.38\paperwidth]{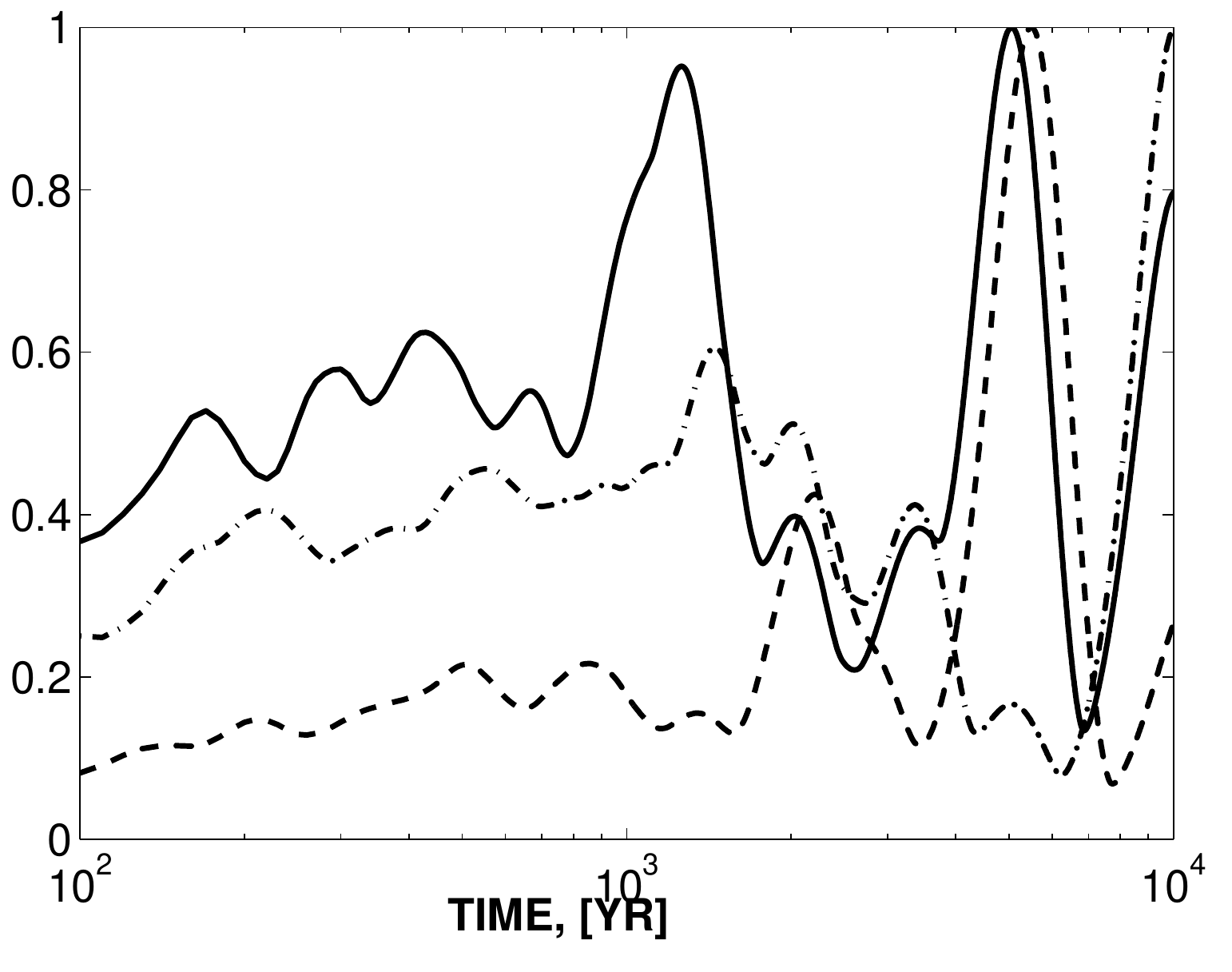}
\par\end{centering}
\caption{Wavelet spectra of the simulated and observational sunspot number
  data sets. The left panel corresponds to short time scales of up to 300 yr. The solid line
  shows the results for the 2D1$\alpha$ model, the dashed line is computed on
  the base of the data set provided by the \cite{hs98} reconstruction and
  the dash-dotted line shows the results for the 2D1$\alpha$L model based on the
  log-normal fluctuations of the $\alpha$-effect. The right panel is similar to the left one, but
  for longer time scales. The dashed line is computed on
  the basis of the  reconstruction provided by \cite{setal04}. The
  spectra were normalized relative to their maxima for clarity.}\label{wavel}
\end{figure*}

Figure~\ref{wavel} shows the global wavelet power spectra
for the sunspot data and for the model data
sets. {The reader can find the
definition of the concept of wavelet power spectra and its
discussion in the context of the solar activity studies in, e.g.,
Frick et al. (1997a). The data were processed  with the standard MATLAB wavelet
toolbox using the standard Morlet wavelet basis. The global
wavelet power spectra were obtained by integrating the spectra in the
time domain (see Eq.(10) in the cited paper). Each spectrum was normalized
relative to its maximum magnitude.}
To illustrate the role of the $\alpha$-effect
fluctuation statistics we show the results for the model with the
log-normal noise $\xi_{\alpha}$. The mean and variance of the
log-normal noise $\xi_{\alpha}$ corresponds to the mean and variance
of the Gaussian noise $\xi_{\alpha}$ in the model 2D1$\alpha$. The
short-term scales spectra look qualitatively similar in all three
sets. The principal difference is the ratio between the spectrum
amplitude for the basic cycle (at 11 years)
 and the amplitude of the second maximum at the period $\approx 200$ years.
This ratio is greater
in the sunspot data set. All three data sets show
the long-term variations with periods of about  100 years, which is usually
identified with the Gleisberg cycle.

The dynamics on the scale of millennia looks similar for all three
data sets, see Figure~\ref{wavel} (right). The model with the
log-normal noise does not show the ordered long-term variations on
this time scale, while the model 2D1$\alpha$ and the reconstruction data
show evidence for the variations with period about $6000$ years.
It is unclear, however, if this result is statistically stable.

\begin{figure}
\begin{centering}
\includegraphics[width=0.98\columnwidth]{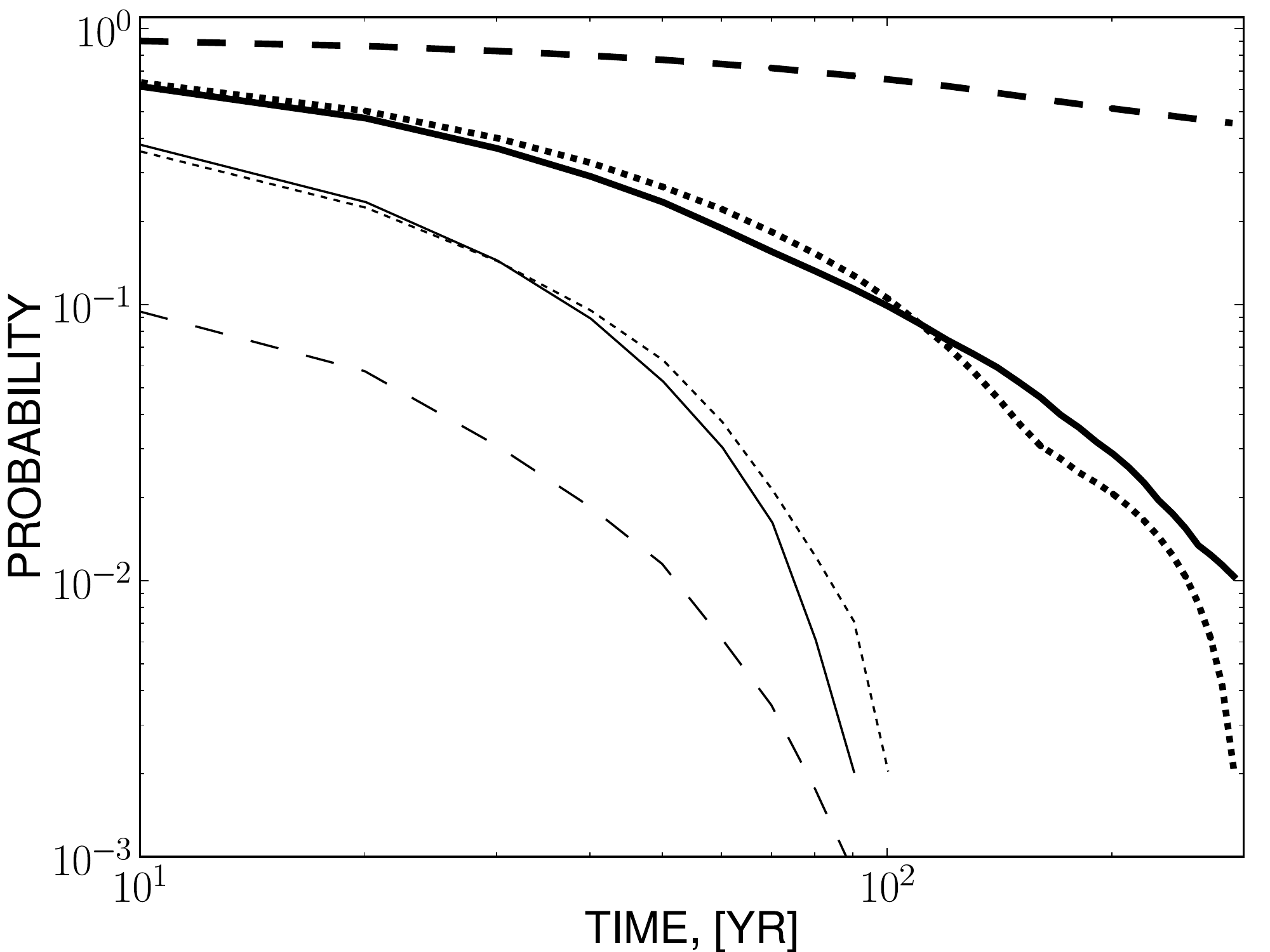}
\par\end{centering}
\caption{Probability of the high- and low-activity 
  episodes for the given duration (see definition in the text). The
  thin lines show the results for the high-activity episodes, and the bold lines the low-activity episodes.
The solid lines (thin and bold) show the results for 2D1$\alpha$ model,
 the dotted lines show the results for the 2D1$\alpha$L model and the  dashed
  lines   show the same for the  reconstruction data set provided by \cite{setal04}.\label{prob}}
\end{figure}

An important statistical property of the dynamo is the occurrence probability of 
 high- and low-activity episodes.
Following \cite{setal04}, we
defined the high-activity episode as having { the average SN $\ge$ 50 (the
minimal averaged SN was higher than 50, $\min \langle(SN)\rangle \ge 50$) and similar for the low-activity
episode occurs when $\max\langle(SN)\rangle \le 50$}.
Then,  we counted the number of episodes for each time scale with
the high and low activity episodes and computed  the probability  distributions
as a function of the time scale. The results are shown in
Figure~\ref{prob}. We find that the dynamo models show a somewhat higher
probability for the high-activity
episodes than the reconstruction data set. Their probability profiles
looks similar in all three cases and   show a significant drop
in the pass from decadal to the centennial time scale. {The probability
of a high-activity episode to occur decreases exponentially with time.
The probability of the opposite event, i.e., a minimum with the average SSN below 50,
increases accordingly.
Note that the end of a high-activity episode does not necessarily imply a low-activity
 epoch.
It could be a local minimum with a duration of less than 20 years (the 11-year interval was used for averaging).}
A similar behavior is found for the probability of the low-activity
episode in the
dynamo models that show a
significant drop of the probability profiles around half-millennium.
The reconstruction data set is very different in this aspect. It seems
that it is not possible to explain all the basic properties of the
sunspot cycle variations as the result of fluctuations of the
$\alpha$-effect.

\begin{figure}
\begin{centering}
\includegraphics[width=0.98\columnwidth]{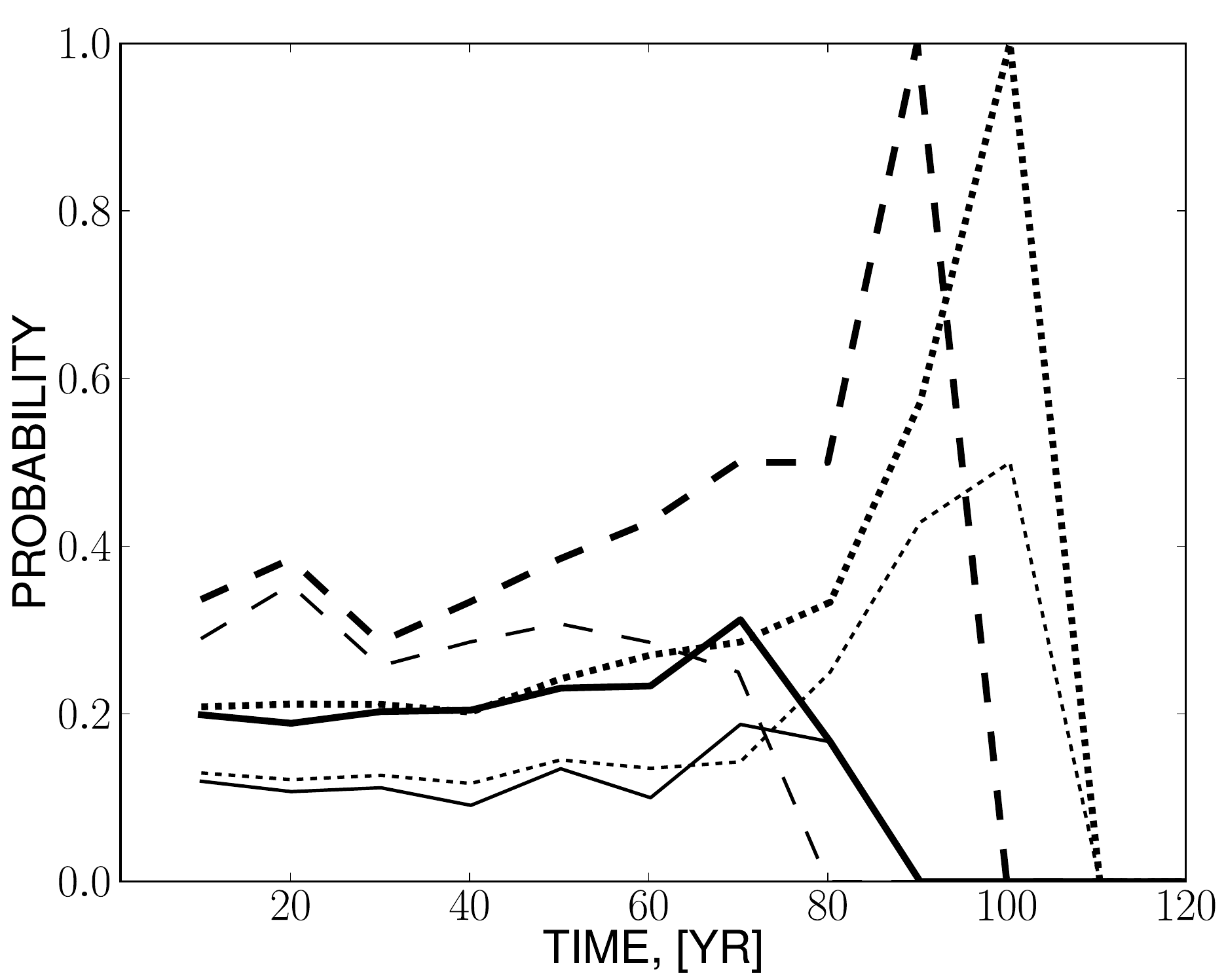}
\par\end{centering}
\caption{Probability that the high-activity episode will be ended
  by low-activity episode of a given length, 30 years length - bold
  lines and 50 years length - thin lines. The solid lines show the results for  2D1$\alpha$ model,
 the dotted lines show the results for the 2D1$\alpha$L model based on the
  log-normal fluctuations of the $\alpha$-effect, and the  dashed lines
  show the results for the  reconstruction provided by \cite{setal04}.\label{prob2}}
\end{figure}

Can one predict that the high-activity episode will be ended with the low-activity episode of the
given length?  Figure \ref{prob2} shows the results for
the low-activity episodes of 30 and 50 years. The
given estimates can be biased because of  the unpresentable
statistics for these  events.

\section{Discussion and Conclusions}
We have studied the impact of low-amplitude Gaussian fluctuations of the
$\alpha$-effect on the statistical properties of the magnetic dynamo
cycle, such as the Waldmeier
relations and the Gnevyshev-Ohl rules. The dynamo model includes long-term fluctuations
of the $\alpha$-effect and employs two types of a
nonlinear feedback of the mean-field on the $\alpha$-effect, including
 algebraic quenching and  dynamic quenching due to the magnetic
helicity generation. The general properties of the dynamo, such as the
direction of the toroidal magnetic field drift, the polar magnetic
field sign reversal at the maximum of a cycle, etc., are consistent
with observations.

Our model does not include the meridional circulation effect,
 which is advocated by the Babcock-Leighton and the
flux-transport type dynamo models
\citep[e.g.,][]{choud95,dc99,detal08}. It was shown that a part of the
Waldmeier relations can be possibly explained by a specially tuned
flux-transport model that considers fluctuations of the
meridional circulation speed \citep{karak11}. Still, observational
constraints on the distribution of the meridional circulation inside
the convection zone are not  very strong because we have measurements
for the surface.  The angular momentum
balance in mean-field models supports for the circulation pattern, which has a deep
circulation stagnation point and a strong
concentration of the velocity speed towards the bottom and the top boundaries
of the solar convection zone (e.g., \citealp{kit11}). Yet, most of the
flux-transport models (including \citealp{karak11}) use a very different
circulation pattern. Following to this argumentation, we postpone a
more complete study of the effects of the meridional circulation
fluctuations to the future.

We showed, confirming the previous findings of \cite{pk11} and \cite{ps11}, that
variations of the $\alpha$-effect amplitude result in variations of the
cycle amplitude and period. Taking into account random fluctuations  of the
$\alpha$-effect, we calculated statistical properties
relating the cycle amplitude, the cycle shape, the rise
time, etc., on the basis of the simulated SN data set covering 
period of more than 10000 years. Our results agree well  with
observations for the Waldmeier
relations and the Gnevyshev-Ohl rules.

From the qualitative point of
view these results were
anticipated from the earlier analysis of the helicity fluctuation effect
in the dynamo given by \cite{choud92} and \cite{h93}
 (see, also \citealp{oss-h96a,oss-h96b,moss-sok08,uetal09}). Our
 results presented in Figure~\ref{minsurf} about the correlation of
 the polar dipole field and the cycle amplitude and as the results for
 the Gnevyshev-Ohl rules suggest that the Waldmeier relations can be
 understood by considering the general properties of the
 magnetic field generation processes, which are involved in the
 dynamo.

Our model shows a good correlation (with low variance) between
the strength of the polar dipole magnetic
field in the cycle minimum and the amplitude of the subsequent cycle.
 This results from the deterministic process of the toroidal
magnetic field generation by the differential rotation from the
large-scale poloidal magnetic field. This correlation is often used
for the cycle prediction \citep{hath09} by Babcock-Leighton type
dynamo models and it is for the first time demonstrated in the
mean-field dynamo. The rise rate of the sunspot
cycle depends on the differential rotation and the amplitude of
the poloidal field. Therefore, the correlation between the rise rate and
amplitude of the cycle is a derivative property and is a
consequence  of the link  between the polar dipole magnetic
field in the cycle minimum and the strength of the toroidal field in
 the subsequent cycle.

Furthermore, following the general idea of \cite{GM78} (cf,
\citealp{h93,chetal07}),
 we can interpret  the  Gnevyshev-Ohl rule as evidence
that the solar cycle is a nonlinear self-excited oscillation that
tends to preserve the property of the attractor under random
perturbations. The amplitude and phase of the subsequent cycles are
related by the so-called Zaslavsky map. The strength of the link between the
parameters of the subsequent cycles is controlled by the fluctuation amplitude
and by the perturbation's decrement. The latter strongly depends on the
nonlinear mechanisms involved in the dynamo. To examine this idea we
made additional simulations with a lower helicity dissipation
rate (high $R_{\chi}$) and found that the correlation coefficients
between the parameters of the subsequent cycle increase with increasing
$R_{\chi}$. Therefore  the link between the odd and
even cycles, and  the period to amplitude
correlation in subsequent cycles can be considered  as evidence
for the fluctuation impact on the dynamo and evidence for nonlinear damping of
these perturbations in the dynamo. This conclusion needs to be
investigated in more detail especially by comparing the results of the
$\alpha$-effect and meridional circulation fluctuations.

\begin{figure}
\begin{centering}
\includegraphics[width=0.95\columnwidth]{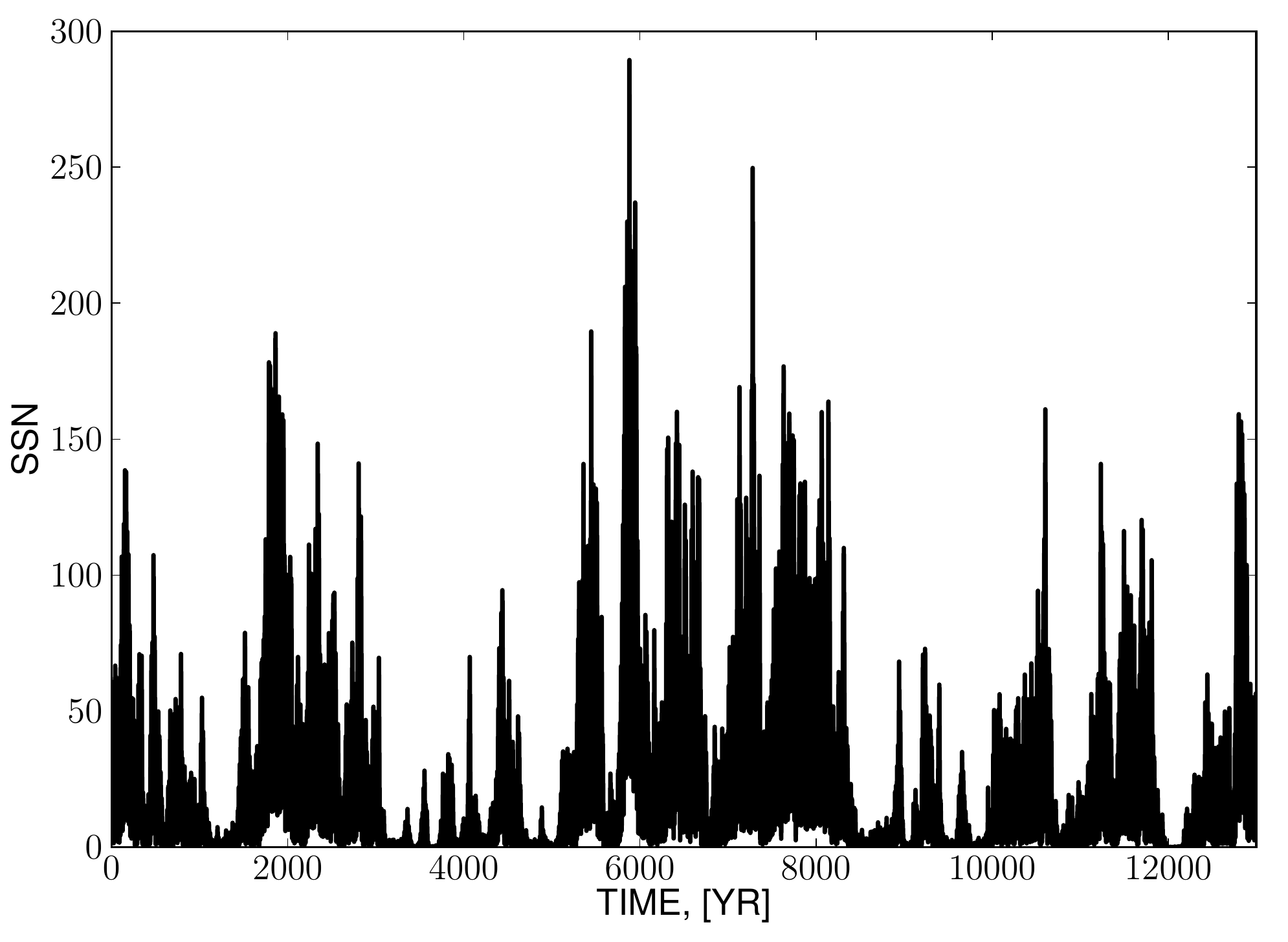}
\par\end{centering}
\caption{Time series of the simulated sunspot number for the
  extended 2D1$\alpha$ model involving the fluctuations of the alpha effect and the
  magnetic feedback on the differential rotation.}\label{ndifr}
\end{figure}

Long-term variations of the magnetic cycle in the dynamo can be
induced in different ways. Two main mechanisms can be identified:
 nonlinear deterministic chaos and an effect of
fluctuations of the turbulent parameters involved in the dynamo
process. Generally, we anticipate that statistical properties of
 long-term cycle variations are depended on the force that
drives the long-term variations. We examined weakly
nonlinear models with the amplitude of the $\alpha$-effect close
to the threshold. In our models, the typical ratio between the
energy of the large-scale toroidal field and the kinetic energy of
 convective flows does not exceed $0.3$. As a result, the chaotic
regime in the model is not as evident as the impact of the
$\alpha$-effect fluctuations.  Figure~\ref{prob} illustrates the difficulty
 to obtain extended episodes of low magnetic
activity in this case, while these episodes are common in the solar
dynamo \citep{setal04}. To amplify the chaotic regime in the model, we
have tried additional possibilities and included an angular
 momentum balance into the dynamo problem to take into account the nonlinear feedback of
the magnetic field on the differential rotation. The model was
described earlier by \cite{p99,p04}.
Figure~\ref{ndifr} shows the simulated SN for the model involving the
nonlinear effect of the magnetic field on the angular momentum balance
in the solar convection zone. This model shows higher intermittency
in the cycle variations than that in Figure~\ref{waldr}, and indeed
it has a similar probability of the occurrence of lowactivity
episodes as that in the reconstruction data set.

It is natural to expect that at least  stellar magnetic cycles of
solar-like stars should demonstrate a variability similar to the
solar one, including relations comparable with the Waldmeier
relations. Available observations of stellar activity provide some
hints that support this expectation.  { Stellar activity data of the
Mount Wilson HK project measuring the Ca H and K line index for
other stars \citep{bal} are available for two activity cycles.} The
wavelet analysis of the data for several stars \citep{balfrick}
demonstrated that the subsequent cycles for a given star can differ
in their cycle amplitudes. We note that a monitoring of
stellar activity of solar-like stars to obtain relations
similar to the Waldmeier ones could establish our prognostic abilities of
solar activity based on these relations much better.

Summarizing the results of the paper, we conclude that the mean-field
solar dynamo theory provides  a way to explain the cycle-to-cycle
variability of solar activity as recorded in sunspots. The results
given in the literature and the results obtained in the paper
suggest that the Waldmeier relations can be explained invoking very
different kinds of dynamo models. More work is necessary to
study the relations between the statistical properties of the dynamo
cycle, and the dynamo mechanisms involved in the magnetic activity
will help to obtain more insight into the processes operating in
the stellar and solar dynamo.
\begin{acknowledgements}
The authors thanks the anonymous referee for the helpful suggestions and
comments. D.S. and V.P. thank for the support the RFBR grant
12-02-00170-a.  Also, V.P. thanks  for  the support of the  Integration Project of SB RAS
\textnumero{34}, the RFBR grant 10-02-00148-a and the partial support
 by the state contracts 02.740.11.0576, 16.518.11.7065 of the Ministry
 of Education and Science of Russian Federation.
\end{acknowledgements}


\begin{appendix}
\section{Appendix }
Here, we describe the contributions of the mean-electromotive force that
are involved in Eq.(\ref{eq:EMF-1}). The basic
formulation is given in \cite{pi08Gafd}(P08). In this paper we reformulate tensor
$\alpha_{i,j}^{(H)}$, which represents the hydrodynamical part of
the $\alpha$-effect, by using Eq.(23) from P08 in the following form,
\begin{eqnarray}
\alpha_{ij}^{(H)} & = & \delta_{ij}\left\{ 3\eta_{T}\left(f_{10}^{(a)}\left(\mathbf{e}\cdot\boldsymbol{\Lambda}^{(\rho)}\right)+f_{11}^{(a)}\left(\mathbf{e}\cdot\boldsymbol{\Lambda}^{(u)}\right)\right)\right\} \label{eq:alpha}\\
 & + & e_{i}e_{j}\left\{ 3\eta_{T}\left(f_{5}^{(a)}\left(\mathbf{e}\cdot\boldsymbol{\Lambda}^{(\rho)}\right)+f_{4}^{(a)}\left(\mathbf{e}\cdot\boldsymbol{\Lambda}^{(u)}\right)\right)\right\} \nonumber \\
 & + & 3\eta_{T}\left\{
   \left(e_{i}\Lambda_{j}^{(\rho)}+e_{j}\Lambda_{i}^{(\rho)}\right)f_{6}^{(a)}\right. \nonumber
   \\
&+&\left. \left(e_{i}\Lambda_{j}^{(u)}+e_{j}\Lambda_{i}^{(u)}\right)f_{8}^{(a)}\right\} .\nonumber
\end{eqnarray}
 The contribution of magnetic helicity $\overline{\chi}=\overline{\mathbf{a\cdot}\mathbf{b}}$
($\mathbf{a}$ is a fluctuating vector magnetic field potential) to
the $\alpha$-effect is defined as $\alpha_{ij}^{(M)}=C_{ij}^{(\chi)}\overline{\chi}$,
where
\begin{equation}
C_{ij}^{(\chi)}=2f_{2}^{(a)}\delta_{ij}\frac{\tau_{c}}{\mu_{0}\overline{\rho}\ell^{2}}-2f_{1}^{(a)}e_{i}e_{j}\frac{\tau_{c}}{\mu_{0}\overline{\rho}\ell^{2}}.\label{alpM}
\end{equation}
 The turbulent pumping, $\gamma_{i,j}$, is also part of the mean
electromotive force in Eq.(23)(P08). Here we rewrite it in a more
traditional form (cf, e.g., ),
\begin{eqnarray}
\gamma_{ij}& = &3\eta_{T}\left\{
  f_{3}^{(a)}\Lambda_{n}^{(\rho)}+f_{1}^{(a)}\left(\mathbf{e}\cdot\boldsymbol{\Lambda}^{(\rho)}\right)e_{n}\right\}
\varepsilon_{inj}\label{eq:pump}\\
&-& 3\eta_{T}f_{1}^{(a)}e_{j}\varepsilon_{inm}e_{n}\Lambda_{m}^{(\rho)}\nonumber
\end{eqnarray}
 The effect of turbulent diffusivity, which is anisotropic because of
the Coriolis force, is given by
\begin{equation}
\eta_{ijk}=3\eta_{T}\left\{
  \left(2f_{1}^{(a)}-f_{1}^{(d)}\right)\varepsilon_{ijk}
-2f_{1}^{(a)}e_{i}e_{n}\varepsilon_{njk}\right\} .\label{eq:diff}
\end{equation}
 Functions $f_{\{1-11\}}^{(a,d)}$ depend on the Coriolis number $\Omega^{*}=2\tau_{c}\Omega_{0}$
and the typical convective turnover time in the mixing-length approximation,
$\tau_{c}=\ell/u'$. They can be found in P08. The turbulent diffusivity
is parametrized in the form, $\eta_{T}=C_{\eta}\eta_{T}^{(0)}$, where
$\eta_{T}^{(0)}={\displaystyle \frac{u'\ell}{3}}$ is the characteristic
mixing-length turbulent diffusivity, $u'$ is the RMS convective velocity,
$\ell$ is the mixing length, $C_{\eta}$ is a constant to control
the intensity of turbulent mixing. The other quantities in Eqs.(\ref{eq:alpha},\ref{eq:pump},\ref{eq:diff})
are $\mathbf{\boldsymbol{\Lambda}}^{(\rho)}=\boldsymbol{\nabla}\log\overline{\rho}$
is the density stratification scale, $\mathbf{\boldsymbol{\Lambda}}^{(u)}=\boldsymbol{\nabla}\log\left(\eta_{T}^{(0)}\right)$
is the scale of turbulent diffusivity, $\mathbf{e}=\boldsymbol{\Omega}/\left|\Omega\right|$
is a unit vector along the axis of rotation. Equations (\ref{eq:alpha},\ref{eq:pump},\ref{eq:diff})
take into account the influence of the fluctuating small-scale magnetic
fields, which can be present in the background turbulence and stem
from the small-scale dynamo. In our paper, the
parameter $\varepsilon={\displaystyle \frac{\overline{\mathbf{b}^{2}}}{\mu_{0}\overline{\rho}\overline{\mathbf{u}^{2}}}}$,
which measures the ratio between the magnetic and kinetic energies
of fluctuations in the background turbulence, is assumed to be equal to
1. This corresponds to the energy equipartition. The quenching function
of the hydrodynamical part of $\alpha$-effect is defined by
\begin{equation}
\psi_{\alpha}=\frac{5}{128\beta^{4}}\left(16\beta^{2}-3-3\left(4\beta^{2}-1\right)\frac{\arctan\left(2\beta\right)}{2\beta}\right).
\end{equation}
 Note in the notation of P08 $\psi_{\alpha}=-3/4\phi_{6}^{(a)}$, and
$\beta={\displaystyle \frac{\left|\overline{B}\right|}{u'\sqrt{\mu_{0}\overline{\rho}}}}$.

Below we give the functions of the Coriolis number defining the dependence
of the turbulent transport generation and diffusivities on the angular
velocity:
\begin{eqnarray*}
f_{1}^{(a)} & = & \frac{1}{4\Omega^{*\,2}}\left(\left(\Omega^{*\,2}+3\right)\frac{\arctan\Omega^{*}}{\Omega^{*}}-3\right),\\
f_{2}^{(a)} & = & \frac{1}{4\Omega^{*\,2}}\left(\left(\Omega^{*\,2}+1\right)\frac{\arctan\Omega^{*}}{\Omega^{*}}-1\right),\\
f_{3}^{(a)} & = & \frac{1}{4\Omega^{*\,2}}\left(\left(\left(\varepsilon-1\right)\Omega^{*\,2}+\varepsilon-3\right)\frac{\arctan\Omega^{*}}{\Omega^{*}}+3-\varepsilon\right),\\
f_{4}^{(a)} & = &
\frac{1}{6\Omega^{*\,3}}\left(3\left(\Omega^{*4}+6\varepsilon\Omega^{*2}+10\varepsilon-5\right)\frac{\arctan\Omega^{*}}{\Omega^{*}}
\right. \\
&-&\left. \left((8\varepsilon+5)\Omega^{*2}+30\varepsilon-15\right)\right),\\
f_{5}^{(a)} & = &
\frac{1}{3\Omega^{*\,3}}\left(3\left(\Omega^{*4}+3\varepsilon\Omega^{*2}+5(\varepsilon-1)\right)\frac{\arctan\Omega^{*}}{\Omega^{*}}
  \right. \\
&-&\left. \left((4\varepsilon+5)\Omega^{*2}+15(\varepsilon-1)\right)\right),\\
f_{6}^{(a)} & = &
-\frac{1}{48\Omega^{*\,3}}\left(3\left(\left(3\varepsilon-11\right)\Omega^{*2}+5\varepsilon-21\right)\frac{\arctan\Omega^{*}}{\Omega^{*}}\right. \\
&-&\left. \left(4\left(\varepsilon-3\right)\Omega^{*2}+15\varepsilon-63\right)\right),\\
f_{8}^{(a)} & = &
-\frac{1}{12\Omega^{*\,3}}\left(3\left(\left(3\varepsilon+1\right)\Omega^{*2}+4\varepsilon-2\right)\frac{\arctan\Omega^{*}}{\Omega^{*}}\right.\\
&-&\left. \left(5\left(\varepsilon+1\right)\Omega^{*2}+12\varepsilon-6\right)\right),\\
f_{10}^{(a)} & = &
-\frac{1}{3\Omega^{*\,3}}\left(3\left(\Omega^{*2}+1\right)\left(\Omega^{*2}+\varepsilon-1\right)\frac{\arctan\Omega^{*}}{\Omega^{*}}\right. \\
&-&\left. \left(\left(2\varepsilon+1\right)\Omega^{*2}+3\varepsilon-3\right)\right),\\
f_{11}^{(a)} & = &
-\frac{1}{6\Omega^{*\,3}}\left(3\left(\Omega^{*2}+1\right)\left(\Omega^{*2}+2\varepsilon-1\right)\frac{\arctan\Omega^{*}}{\Omega^{*}}\right. \\
&-&\left. \left(\left(4\varepsilon+1\right)\Omega^{*2}+6\varepsilon-3\right)\right).\\
f_{2}^{(d)} & = &
 \frac{1}{4\Omega^{*\,2}}\left(\left(\left(\varepsilon-1\right)\Omega^{*\,2}+3\varepsilon+1\right)\frac{\arctan\Omega^{*}}{\Omega^{*}}-3\varepsilon-1\right).\\
\end{eqnarray*}
\end{appendix}

\end{document}